\documentclass[12pt,a4paper]{article}

\usepackage{amssymb}
\usepackage{amsmath}
\usepackage[dvips]{graphicx,psfrag}
\usepackage{epsfig}
 
\newcommand{\be}{\begin{equation}}
\newcommand{\ee}{\end{equation}}
\newcommand{\bea}{\begin{eqnarray}}
\newcommand{\eea}{\end{eqnarray}}
\newcommand{\lb}{\label}
\newcommand{\bdm}{\begin{displaymath}}
\newcommand{\edm}{\end{displaymath}}
\newcommand{\D}{{\rm d}}
\newcommand{\E}{{\rm e}}
\newcommand{\I}{{\rm i}}
\newcommand{\X}{{\mathbf x}}

\begin{document}

\begin{titlepage}

\noindent
\begin{center}
\vspace*{1cm}

{\bf\textsc CONCEPTUAL PROBLEMS IN QUANTUM GRAVITY \\ AND QUANTUM
  COSMOLOGY}   
  
\vskip 1cm

{\bf Claus Kiefer} 
\vskip 0.4cm
Institute for Theoretical Physics,\\ University of Cologne, \\
Z\"ulpicher Strasse~77,
50937 K\"oln, Germany.\\ {\tt http://www.thp.uni-koeln.de/gravitation/}
\vspace{1cm}

\begin{abstract}
The search for a consistent and empirically established quantum theory
of gravity is among the biggest open problems of fundamental
physics. The obstacles are of formal and of conceptual nature. Here, I
address the main conceptual problems, discuss their present
status and outline further directions of research. For this purpose,
the main current approaches to quantum 
gravity are briefly reviewed and compared. 
\end{abstract}

\vskip 1cm

{\bf Journal-ref.: ISRN Math.Phys. 2013 (2013) 509316}

\end{center}

\end{titlepage}


\section{Quantum theory and gravity -- what is the connection?}

According to our current knowledge, the fundamental interactions of
Nature are the strong, the electromagnetic, the weak, and the
gravitational interactions. The first three are successfully described
by the Standard Model of particle physics, in which a partial
unification of the electromagnetic and the weak interactions has been
achieved. Except for the non-vanishing neutrino masses, there
exist at present no empirical fact that is clearly at variance with the
Standard Model. Gravity is described by Einstein's theory of general
relativity (GR), and no empirical fact is known that is in clear
contradiction to GR. From a pure empirical point of view, we thus have no
reason to search for new physical laws. From a theoretical
(mathematical and conceptual) point of view, however, the situation is
not satisfactory. Whereas the Standard Model is a quantum field theory
describing an incomplete unification of interactions, GR is a
classical theory. Let us have a brief look at Einstein's theory, see,
for example, Misner {\em et al.} (1973). 
It can be defined by the Einstein--Hilbert action
\be
S_{\rm EH}= \frac{c^4}{16\pi G}\int_{\mathcal M}{\rm d}^4x\ \sqrt{-g}
 \left(R-2\Lambda\right)- \frac{c^4}{8\pi G}
\int_{\partial\mathcal M}{\rm d}^3x\ \sqrt{h}K  ,
\lb{EH}
\ee
where $g$ is the determinant of the metric, $R$ the Ricci scalar, and
$\Lambda$ is the cosmological constant. In addition to the two main terms,
which consist of integrals over a spacetime region ${\mathcal M}$,
there is a term that is defined on the boundary ${\partial\mathcal M}$
(here assumed to be space-like) of this region. This term is needed
for a consistent 
variational principle; here, $h$ is the determinant of the
three-dimensional metric, and $K$ is the trace of the second
fundamental form. 

In the presence of non-gravitational fields, \eqref{EH} is augmented
by a `matter action' $S_{\rm m}$. From the sum of these actions,
one finds Einstein's field equations by variation with respect to the metric, 
\be
G_{\mu\nu}:= R_{\mu\nu}-\frac{1}{2}g_{\mu\nu}R
= \frac{8\pi G}{c^4}T_{\mu\nu}-\Lambda g_{\mu\nu} .
\lb{einstein}
\ee
The right-hand side displays the symmetric (Belinfante)
energy--momentum tensor
\be
T_{\mu\nu}=\frac{2}{\sqrt{-g}}\frac{\delta S_{\rm m}}{\delta g^{\mu\nu}} ,
\lb{em}
\ee
plus the cosmological-constant term, which may itself be accommodated
into the energy--momentum tensor as a contribution of the `vacuum energy'. 
If fermionic fields are added, one must generalize GR to the
Einstein--Cartan theory or to the Poincar\'e gauge theory, because spin
is the source of torsion, a geometric quantity that is identically
zero in GR (see e.g. Gronwald and Hehl 1996).

As one recognizes from \eqref{einstein}, these equations can no longer
have exactly the same form if the quantum nature of the fields in
$T_{\mu\nu}$ is taken into account. For then we have operators in
Hilbert space on the right-hand side and classical functions on the
left-hand side. A straightforward generalization would be to replace
$T_{\mu\nu}$ by its quantum expectation value, 
\be
\lb{semi}
R_{\mu\nu}-\frac{1}{2}g_{\mu\nu}R+\Lambda g_{\mu\nu}=
  \frac{8\pi G}{c^4}\langle\Psi\vert\hat{T}_{\mu\nu}
\vert\Psi\rangle .
\ee
These `semiclassical Einstein equations' lead to problems when viewed
as exact equations at the most fundamental level,
cf. Carlip (2008) and the references therein. They spoil the linearity of
quantum theory and even seem to be in
conflict with a performed experiment (Page and Geilker~1981). They
may nevertheless be of some value in an approximate way. Independent
of the problems with \eqref{semi}, one can try to test them in a
simple setting such as the Schr\"odinger--Newton equation; it seems,
however, that such a test is not realisable in the foreseeable future
(Giulini and Grossardt 2011). 
This poses the question of the
connection between gravity and quantum theory (Kiefer 2012).

Despite its name, quantum theory is not a particular theory for a
particular interaction. It is rather a general framework for physical
theories, whose fundamental concepts have so far exhibited an amazing
universality. Despite the ongoing discussion about its
interpretational foundations (which we shall address in the last
section), the concepts of states in Hilbert space, and in particular
the superposition principle, have successfully passed thousands of
experimental tests.

It is, in fact, the superposition principle that points towards the
need for quantizing gravity. In the 1957 Chapel Hill Conference,
Richard Feynman gave the following argument (DeWitt and Rickles~2011,
pp.~250--60), see also Zeh (2011). He considers a Stern--Gerlach type
of experiment in 
which two spin-1/2 particles are put into a superposition of spin up
and spin down and is guided to two counters. He then imagines a connection of
the counters to a ball of macroscopic dimensions. The superposition of
the particles is thereby transferred to a superposition of the ball
being simultaneously at two positions. But this means that the ball's
gravitational field is in a superposition, too! In Feynman's own words
(DeWitt and Rickles~2011, p.~251):

\begin{quote}
Now, how do we analyze this experiment according to quantum mechanics?
We have an amplitude that the ball is up, and an amplitude that the
ball is down. That is, we have an amplitude (from a wave function)
that the spin of the electron in the first part of the equipment is
either up or down. And if we imagine that the ball can be analyzed
through the interconnections up to this dimension ($\approx 1$ cm) by
the quantum mechanics, then before we make an observation we still
have to give an amplitude that the ball is up and an amplitude that
the ball is down. Now, since the ball is big enough to produce a {\em
  real} gravitational field \ldots we could use that gravitational
field to move another ball, and amplify that, and use the connections
to the second ball as the measuring equipment. We would then have to
analyze through the channel provided by the gravitational field itself
via the quantum mechanical amplitudes.
\end{quote}

In other words, the gravitational field must then be described by
quantum states subject to the superposition principle. The same
argument can, of course, be applied to the electromagnetic field (for
which we have strong empirical support that it is of quantum nature). 

Feynman's argument is, of course, not an argument of logic that would
demand the quantization of gravity with necessity. It is an argument
based on conservative heuristic ideas that proceed from the
extrapolation of established and empirically confirmed concepts (here,
the superposition principle) beyond their present range of
application. It is in this way that physics usually evolves. 
Albers {\em et al.} (2008) contains a detailed account of arguments
that demand the necessity of quantizing the gravitational field. It is
shown that all these arguments are of a heuristic value and that they do not
lead to the quantization of gravity by a logical conclusion. It is
conceivable, for example, that the linearity of quantum theory breaks
down in situations where the gravitational field becomes strong, see,
for example, Penrose (1996), Singh (2005), and Bassi {\em et al.}
(2013). But one has to emphasize 
that no empirical hint for such a drastic modification exists so far. 

Alternatives to the direct quantization of the gravitational field
include what is called `emergent gravity', cf. Padmanabhan (2010) and
the references therein. Motivated by the thermodynamic properties of
black holes (see below), one might get the impression that the
gravitational field is an effective thermodynamic entity that does
not demand its direct quantization, but points to the existence of
new, so far unknown microscopic degrees of freedom underlying
gravity. Even if this were the case (which is far from clear), there
may exist microscopic degrees of freedom for which quantum theory
would apply. Whether this leads to a quantized metric or not, is not
clear. In string theory (see below), gravity is an
emergent interaction, but still the metric is quantized, and the standard
approach of quantum 
gravitational perturbation theory is naturally implemented.  
In the following, we restrict ourselves to approaches in which a
quantized metric makes sense.

Besides the general argument put forward by Feynman, there exist a
couple of further arguments that suggest that the gravitational interaction
be quantized (Kiefer~2012). Let me briefly review three of them.

The first motivation comes from the continuation of the reductionist
programme. In physics, the idea of unification has been very
successful, culminating so far in the Standard Model of strong and
electroweak interactions. Gravity acts universally to all form of
energies. A unified theory of {\em all} interactions including gravity
should thus not be a hybrid theory in using classical and quantum
concepts. A coherent quantum theory of all interactions (often called
`theory of everything' or TOE) should thus also include a quantum
description of the gravitational field. 

A second motivation is the unavoidable presence of singularities in
Einstein's theory of GR , see, for example Hawking and Penrose (1996)
and Rendall (2005). Prominent examples are the cases of the big bang
and the interior of black holes. 
One would thus expect that a more fundamental
theory encompassing GR does not predict any singularities. This would
be similar to the case of the classical singularities from
electrodynamics, which are avoided in quantum electrodynamics.
The fate of
the classical singularities in some approaches to quantum gravity 
will be discussed below.

A third motivation is known as the `problem of time'. In quantum
mechanics, time is absolute. The parameter $t$ occurring in the
Schr\"odinger equation has been directly inherited from Newtonian
mechanics and is not turned into an operator. In quantum field theory,
time by itself is no longer absolute, but the four-dimensional
spacetime is; it constitutes the fixed background structure on which
the dynamical fields act. GR is of a very different nature. According
to the Einstein equations \eqref{einstein}, spacetime is dynamical,
acting in a complicated manner with energy--momentum of matter and
with itself. The concepts of time (spacetime) in quantum theory and GR
are thus drastically different and cannot both be fundamentally
true. One thus needs a more fundamental theory with a coherent notion
of time. The absence of a fixed background structure is also called
`background independence' (cf. Anderson 1967) and often used a
leitmotiv in the search for 
a quantum theory of gravity, although it is not quite the same as the
problem of time, as we shall see below.

A central problem in the search for a quantum theory of gravity is the
current lack of a clear empirical guideline. This is partly related to
the fact that the relevant scales, on which quantum effects of gravity
should definitely be relevant, is far remote from being directly
explorable. The scale is referred to as the Planck scale and consists of
the Planck length, $l_{\rm P}$, Planck time, $t_{\rm P}$,
 and Planck mass, $m_{\rm P}$, respectively. They are given by the expressions
\bea 
l_{\rm P} &:=& \sqrt{\frac{\hbar G}{c^3}} \approx 1.62\times 10^{-33}\ 
{\rm cm} ,
\lb{lP}\\
t_{\rm P} &:=& \frac{l_{\rm P}}{c}=\sqrt{\frac{\hbar G}{c^5}}
\approx 5.39\times 10^{-44}\ {\rm s} ,\\
\lb{tP}
m_{\rm P} &:=& \frac{\hbar}{l_{\rm P}c}=\sqrt{\frac{\hbar c}{G}}
\approx 2.18\times 10^{-5}\ {\rm g}\approx 1.22 \times 10^{19}\ {\rm GeV}/c^2 .
\lb{mP}
\eea 
It must be emphasized that units of length, time, and mass cannot be
formed out of $G$ and $c$ (GR) or out of 
$\hbar$ and $c$ (quantum theory) alone. 

In view of the Planck scale, the Standard Model provides an additional
motivation for quantum gravity:
it seems that the Standard Model does not exist as a consistent quantum
field theory up to arbitrarily high energies, see, for example,
Nicolai (2013). The reasons for this
failure may be a potential instability of the effective potential
and the existence of Landau poles. The Standard Model can thus by itself not
be a fundamental theory, although with the recently measured
Higgs mass $m_{\rm H}$ of about 126 GeV it is in principle conceivable that it
holds up to the Planck scale. It is so far an open issue why the Higgs
mass is stabilized at such a low value and not driven to high energies
by quantum loop corrections (`hierarchy problem'). In fact, one has
\be
\left(\frac{m_{\rm P}}{m_{\rm H}}\right)^2\sim 10^{34}.
\ee
Possible solutions to the hierarchy problem include supersymmetric
models and models with higher dimensions, but a clear solution is not
yet available.

In astrophysics, the Planck scale is usually of no relevance. The
reason is that structures in the Universe, with the exception of black
holes, occur at (length, time, and mass) scales that are different
from the Planck scale by many orders of magnitude. 
This difference is quantified by the `fine-structure constant of
gravity' defined by
\be
\alpha_{\rm g}:= \frac{Gm_{\rm pr}^2}{\hbar c}=
\left(\frac{m_{\rm pr}}{m_{\rm P}}\right)^2 \approx 5.91\times 10^{-39} ,
\lb{alphag}
\ee
where $m_{\rm pr}$ denotes the proton mass. The Chandrasekhar mass,
for example, which gives the correct order of magnitude for main
sequence stars like the Sun, is given by $\alpha_{\rm g}^{-3/2}m_{\rm
  pr}$. 

Let us have at the end of this section a brief look at the connection
between quantum theory and gravity at the level where gravity is
treated as a classical interaction (Kiefer~2012). The lowest level is 
quantum mechanics plus Newtonian gravity, at which many experimental
tests exist. 
Most of them employ atom or neutron interferometry and
can be described by the Schr\"odinger equation with the
Hamiltonian given by
\be
H=\frac{{\mathbf p}^2}{2m} +m{\mathbf g}{\mathbf r}-
 \mbox{\boldmath$\omega$}{\mathbf L} ,
\lb{HCOW}
\ee
where the second term is the Newtonian potential in the limit of
constant gravitational acceleration, and the last term describes a
coupling between the rotation of the Earth (or another rotating
system) and the angular momentum of the particle. 
An interesting recent suggestion in this context is the possibility to
see the general relativistic time dilatation in the interference
pattern produced by the interference of two partial particle beams at
different heights in the gravitational field
(Zych {\em et al.}~2012). A more general treatment is based on the
Dirac equation and its non-relativistic (`Foldy--Wouthuysen') expansion.

The next level is quantum field theory in a curved spacetime (or in a
flat spacetime, but in non-inertial coordinates). Here, concrete
prediction are available, although they have so far not been
empirically confirmed. The perhaps most famous prediction is the
Hawking effect according to which every stationary black hole is
characterized by the temperature (Hawking~1975)
\be
\lb{TH}
T_{\rm BH}=\frac{\hbar\kappa}{2\pi k_{\rm B}c} ,
\ee
where $\kappa$ is the surface gravity of a stationary black hole, which
by the no-hair theorem is uniquely characterized by its mass $M$,
its angular momentum $J$, and (if present) its electric charge $q$.
In the particular case of the spherically symmetric Schwarzschild 
black hole, one has $\kappa=c^4/4GM=GM/R_{\rm S}^2$, where $R_{\rm
S}=2GM/c^2$ is the Schwarzschild radius, and therefore
\be
\lb{TSchwarz}
T_{\rm BH} =\frac{\hbar c^3}{8\pi k_{\rm B}GM}\approx 6.17\times 10^{-8}
 \left(\frac{M_{\odot}}{M}\right)\ {\rm K} .
\ee
The presence of a temperature for black holes means that these objects
have a finite lifetime. Upon radiating away energy, they become
hotter, releasing even more energy, until all the mass (or almost all
the mass) has been radiated away. For the lifetime
of a Schwarzschild black hole, one finds the expression
(MacGibbon~1991)
\begin{equation}\label{bh-lifetime2}
\tau_\mathrm{BH} \approx 407\left(\frac{f(M_0)}{15.35}\right)^{-1}
\left(\frac{M}{10^{10}\ {\rm g}}\right)^3\ {\rm s}
\approx 6.24\times 10^{-27}M_0^3[{\rm g}]f^{-1}(M_0)\ {\rm s},
\end{equation}
where $f(M_0)$ is a measure of the number of emitted particle species;
it is normalized to
$f(M_0)=1$ for $M_0\gg 10^{17}\ {\rm g}$ when only effectively
massless particles are emitted. If one sums over the contributions
from all particles in the Standard Model up to an energy of about $1\
{\rm TeV}$, one finds $f(M)=15.35$, which motivates the occurrence of
this number in \eqref{bh-lifetime2}. Because the temperature of
 black holes that result 
from stellar collapse is too small to be observable, the hope is that
primordial black holes exist for which the temperature can become high
(Carr 2003).

Since black holes have a temperature, they also have an entropy.
It is called `Bekenstein--Hawking entropy' and is found from
thermodynamic arguments to be given by the universal expression
\begin{equation}\label{SBH}
  S_\mathrm{BH} = \frac{k_\mathrm{B} A}{4 l_{\rm P}^2},
\end{equation}
where $A$ is the surface of the event horizon. 
For the special case of the Schwarzschild black hole, it reads
\begin{equation}\label{SBHSS}
  S_\mathrm{BH} = \frac{k_\mathrm{B} \pi R_{\rm S}^2}{ G \hbar}
  \approx 1.07\times 10^{77}k_\mathrm{B}\left(\frac{M}{M_{\odot}}\right)^2.
\end{equation}
All these expressions depend on the fundamental constants $\hbar$,
$G$, and $c$. They may thus provide a key for quantum
gravity.

In flat spacetime, an effect exists that is analogous to the
black-hole temperature \eqref{TH}. If an observer moves with constant
acceleration through the standard Minkowski vacuum, he will perceive this
state not as empty, but as filled with thermal particles (`Unruh
effect'), see Unruh (1976). The temperature is given by the
`Davies--Unruh temperature' 
\be 
\lb{TDU}
T_{\rm DU}= \frac{\hbar a}{2\pi k_{\rm B}c}\approx 4.05\times 10^{-23}
\ a\left[\frac{\rm cm}{{\rm s}^2}\right]\ {\rm K} .
\ee
The similarity of \eqref{TDU} and \eqref{TH} is connected with the
presence of an event horizon in both cases. A suggestion for an
experimental test of \eqref{TDU} can be found in Thirolf {\em et al.}
(2009). 

Many of the open questions to be addressed
in any theory of quantum gravity are connected with the temperature
and the entropy of black holes (Strominger 2009). 
The two most important questions
concern the microscopic interpretation of entropy
and the final fate of a black hole; the latter is deeply intertwined
with the problem of information loss.

The Bekenstein--Hawking entropy \eqref{SBH} was derived from
thermodynamic considerations, without identifying appropriate
microstates and performing a counting in the sense of statistical
mechanics. Depending on the particular approach to quantum gravity, various ways
of counting states have been developed. In most cases one can get the
behaviour that the statistical entropy $S_{\rm stat}\propto A$,
although not necessarily with the desired factor occurring in
\eqref{SBH}. 

Hawking's calculation leading to \eqref{TH} breaks down when the black
hole becomes small and effects of full quantum gravity are expected to
come into play. This raises the following question. According to
\eqref{TH}, the radiation of the black hole is thermal. What, then,
happens in the final phase of the black-hole evolution? If only
thermal radiation were left, all initial states that lead to a black
hole would end up in one and the same final state -- a thermal state.
That is, the information about the initial state would be lost. This
is certainly in contradiction with standard quantum theory for a
closed system, for which the von Neumann entropy $S=-k_{\rm B}{\rm
  tr}(\rho\ln\rho)$ is constant, where $\rho$ denotes the density
matrix of the system. This possible contradiction is called
 the `information-loss
paradox' or `information-loss problem' for black holes
(Hawking~1976). 

Much has been said since 1976 about the information-loss problem
(see e.g. Page (1994, 2013)), without final consensus. This is not
surprising, because the final solution will only be obtained if the
final theory of quantum gravity is available. Some remarks can 
be made, though. Hawking's original calculation does not show that
black-hole radiation is strictly thermal. It only shows that the expectation
value of the particle-number operator is strictly thermal. There certainly exist
pure quantum states which lead to such a thermal expression (Kiefer
2001, 2004). A relevant example is a two-mode squeezed state, which
after tracing out one mode leads to the density matrix of a canonical
ensemble; such a state can be taken as a model for a quantum state
entangling the interior and exterior of the black hole, see Kiefer
(2001, 2004) and the references therein. 

The same can be said about the Unruh effect. The temperature
\eqref{TDU} can be understood as arising from tracing out degrees of
freedom in the total quantum state, which is pure (Freese {\em et al.}
1985). In the case of a moving mirror, the pure quantum state can
exhibit thermal behaviour, without any information loss (see
e.g. Birrell and Davies 1982, Sec.~4.4). 

For a semiclassical black hole, therefore, the information-loss
problem does not arise. The black hole can, and it fact must, be
treaten as an open quantum system, so that a mixed state emerges from
the process of decoherence (Zeh~2005). The total state of system and
environment (which means everything interacting with the black hole)
can be assumed to stay in a pure state.  Still, there is some
discussion whether something unusual happens at the horizon even for a
large black hole (Almheiri {\em et al.}~2013), but this is a
contentious issue. 

If the black hole approaches the Planck regime, the question about the
information-loss problem is related to the fate of the singularity. 
If the singularity remains in quantum gravity (which is highly
unlikely), information will indeed be destroyed. Most approaches to
quantum gravity indicate that entropy is conserved for the total
system, so there will not be an information-loss problem, in
accordance with standard quantum theory. How the exact quantum state
during the final evaporation phase looks like, is unclear. One can
make oversimplified models with harmonic oscillators (Kiefer {\em et
  al.} 2009), but an exact solution from an approach to quantum
gravity is elusive. 


\section{Main approaches to quantum gravity}

Following Isham (1987), one can divide the approaches roughly into two
classes. 
In the first class, one starts from a given classical theory of
gravity and applies certain quantization rules to arrive at a quantum
theory of gravity. In most cases, the starting point is GR. This does
not yet lead to a unification of interactions; one arrives at a
separate quantum theory for the gravitational field, in analogy to
quantum electrodynamics (QED). Most likely, the resulting theory is an
effective theory only, valid only in certain situations and for
certain scales.\footnote{``It is generally believed today that the
  realistic theories that we use to describe physics at accessible
  energies are what are known `effective field theories'.'' (Weinberg
  1995, p.~499).} Depending on the method used, one distinguishes
between covariant and canonical quantum gravity.

The second class consists of approaches that seek to construct a
unified theory of all interactions. Quantum aspects of gravity are
then seen only in a certain limit -- in the limit where the various
interactions become distinguishable. The main representative of this
second class is string theory. 

In the rest of this section, I shall give a brief overview of the main
approaches. For more details, I refer to Kiefer (2012). In most
expressions, units are chosen with $c=1$.

\subsection{Covariant quantum gravity}

In covariant quantum gravity, one employs methods that make use of
four-dimensional covariance. Today, this is usually done by using the
quantum gravitational path integral, see, for example, Hamber (2009). 
Formally, the path integral reads
\be
\lb{pathgrav}
Z[g]=\int{\mathcal D}g_{\mu\nu}(x)\ \E^{\I S[g_{\mu\nu}(x)]},
\ee
where the sum runs over all metrics on a four-dimensional manifold
${\mathcal M}$ quotiented by the diffeomorphism group
${\rm Diff}{\mathcal M}$. 
In addition, one may wish to perform a sum
over all topologies, because they may also be subject to the
superposition principle. This is, however, not possible in full generality,
because four-manifolds are not classifiable. Considerable care must be
taken in the treatment of the integration measure. 
In order to make it well defined, one has to apply
the Faddeev--Popov procedure known from gauge theories. 

One application of the path integral \eqref{pathgrav} is the
derivation of Feynman rules for the perturbation theory. One makes the ansatz
\be
\lb{2.59}
g_{\mu\nu}=\bar{g}_{\mu\nu}+\sqrt{32\pi G}f_{\mu\nu},
\ee
where $\bar{g}_{\mu\nu}$ denotes the background field with respect to
which covariance is implemented in the
formalism, and $f_{\mu\nu}$ denotes the quantized field (the `gravitons'), with
respect to which the perturbation theory is performed.
Covariance with respect to the background metric means
that no particular background is distinguished; in this sense, `background
independence' is implemented into the formalism.

There is an important difference in the quantum gravitational
perturbation theory as compared to the Standard Model: the theory is
non-re\-norma\-lizable. This means that one encounters a new type of
divergences at each order of perturbation theory, resulting in an
infinite number of free parameters. 
For example, at two loops the following divergence in the Lagrangian
is found (Goroff and Sagnotti~1986) 
\be
\lb{2.79}
{\mathcal L}^{({\rm div})}_{2-{\rm loop}}=
\frac{209\hbar^2}{2880}\frac{32\pi G}{(16\pi^2)^2\epsilon}\sqrt{-\bar{g}}
\bar{R}^{\alpha\beta}_{\ \ \gamma\delta}
\bar{R}^{\gamma\delta}_{\ \ \mu\nu}\bar{R}^{\mu\nu}_{\ \ \alpha\beta} ,
\ee
where $\epsilon=4-D$, with $D$ being the number of spacetime
dimensions, and $\bar{R}^{\mu\nu}_{\ \ \alpha\beta}$ etc. denotes the
Riemann tensor corresponding to the background metric.

New developments have given rise to the hope that a generalization of
covariant quantum general relativity may not even be renormalizable,
but even finite -- this is $N=8$ supergravity. Bern {\em et al.}
(2009) have found that the theory is finite at least up to four
loops. This indicates that a hitherto unknown symmetry may be responsible
for the finiteness of this theory. Whether such a symmetry
really exists and what its nature could be is not known at present.

Independent of the problem of non-renormalizability, one can study
covariant quantum gravity at an effective level, truncating the theory
at, for example, the one-loop level. At this level, concrete
predictions can be made, because the ambiguity from the free
parameters at higher order does not enter. One example is the
calculation of the quantum gravitational correction to the Newtonian
potential between two masses (Bjerrum-Bohr {\em et al.}~2003).
The potential at one-loop order reads
\be
\lb{2.82}
V(r)=-\frac{Gm_1m_2}{r}\left(1+3\frac{G(m_1+m_2)}{rc^2}
+\frac{41}{10\pi}\frac{G\hbar}{r^2c^3}+{\mathcal O}(G^2)\right),
\ee
where the first correction term is an effect from classical GR, and
only the second term is a genuine quantum gravitational correction
term (which is, however, too small to be measurable). 

Apart from perturbation theory, expressions such as \eqref{pathgrav}
for the path integral can, in four spacetime dimensions and higher, only be 
defined and evaluated by numerical methods. One example is Causal
Dynamical Triangulation (CDT) (Ambj\o rn {\em et al.}~2013). Here, spacetime
is foliated into a set of four-dimensional simplices
and Monte Carlo methods are used for the path integral. It was found
that spacetime appears indeed effectively four-dimensional in the
macroscopic limit, but becomes two-dimensonal when approaching the
Planck scale.

This effective two-dimensionality is also seen in another approach of
covariant quantum gravity -- asymptotic safety (see e.g. Nink and
Reuter~2012). A theory is called asymptotically safe if all essential
coupling parameters approach for large energies a fixed point where at
least one of them does not vanish. (If they all vanish, one has the
situation of asymptotic freedom.) Making use of renormalization group
equations, strong indications have been found that quantum GR is
asymptotically safe. If true, this would be an example for a theory of
quantum gravity that is valid at all scales.
The small-scale structure of spacetime is among the most exciting open
problems in quantum gravity, see Carlip (2010) and the articles
collected in the volume edited by Amelino-Camelia and
Kowalski-Glikman (2005).  

\subsection{Canonical approaches}

In canonical approaches to quantum gravity, one constructs a
Hamiltonian formalism at the classical level before quantization. 
In this procedure, spacetime is foliated into a family of spacelike
hypersurfaces. This
leads to the presence of constraints, which are connected with the
invariances of the theory. One has four (local) constraints associated
with the classical diffeomorphisms.  One is the
Hamiltonian constraint ${\mathcal H}_{\perp}$ , which generates
hypersurface deformations 
(many-fingered time evolution); the three other constraints are the momentum or
diffeomorphism constraints ${\mathcal H}_a$, 
which generate three-dimensional coordinate transformations. 
If one uses tetrads instead of
metrics, four additional constraints (`Gauss constraints') associated
with the freedom of performing local Lorentz transformations are
present. Classically, the constraints obey a closed (but not Lie) algebra. 
For the exact relation of the constraints to the classical spacetime
diffeomorphisms, see, for example, Pons {\em et al.} (2009) and
Barbour and Foster (2008).

By quantization, the constraints are turned into quantum constraints
for physically admissible wave functionals. The exact form depends on
the choice of canonical variables. If one uses the three-dimensional
metric as the configuration variable, one arrives at quantum
geometrodynamics. If one uses a certain holonomy as the configuration
variable, one arrives at loop quantum gravity. 

\subsubsection{Quantum geometrodynamics} 

In geometrodynamics, the canonical variables are the three-metric
${h}_{ab}(x)$ and its conjugate momentum ${p}^{cd}(y)$, which is
linearly related to the second fundamental form. In the quantum
theory, they are turned into operators that obey the standard
commutation rules,
\be
\lb{5.3}
[\hat{h}_{ab}(x),\hat{p}^{cd}(y)]=
\I\hbar\delta^c_{(a}\delta^d_{b)}\delta(x,y).
\ee
Adopting a general procedure suggested by Dirac, the constraints are
implemented as quantum constraints on the wave functionals,
\bea
{\mathcal H}_{\perp} \Psi &=& 0, \lb{5.6} \\ 
{\mathcal H}_a \Psi &=& 0. \lb{5.7}
\eea
The first equation is called Wheeler--DeWitt equation
(DeWitt 1967, Wheeler 1968); the three other
equations are called quantum momentum or diffeomorphism
constraints. The latter guarantee that the wave functional
is invariant under three-di\-men\-sio\-nal coordinate
transformations. The configuration space of all three-metrics divided
by three-dimensional diffeomorphisms is called superspace.

In the vacuum case, the above equations assume the explicit form
\bea
\hat{\mathcal H}_{\perp}\Psi&:=&
\left(-16\pi G\hbar^2G_{abcd}\frac{\delta^2}{\delta h_{ab}\delta h_{cd}}
-\frac{\sqrt{h}}{16\pi G}(\,{}^{(3)}\!R-2\Lambda)\right)\Psi=0, \lb{5.21}\\
\hat{\mathcal H}_a\Psi &:=& -2D_bh_{ac}\frac{\hbar}{\I}
\frac{\delta\Psi}{\delta h_{bc}} =0. \lb{5.22}
\eea
Here, $D_b$ is the three-dimensional covariant derivative,
$G_{abcd}$ is the DeWitt metric (which is an ultralocal function of the
three-metric), and ${}^{(3)}\!R$ is the three-dimensional Ricci
scalar. In the presence of non-gravitational fields, the corresponding
pieces of their Hamiltonian are added to the expressions \eqref{5.21}
and \eqref{5.22}. 

There are various problems connected with these equations. One is the
problem to make mathematical sense out of them and to look for
solutions. This includes the implementation of the
positive-definiteness of the three-metric, which may point to the need
of an `affine quantization' (Klauder 2010).
Other problems are of conceptual nature and will be
discussed below. Attempts to derive \eqref{SBH} exist in this
framework, see e.g. Vaz {\em et al.} (2008), but the correct
proportionality factor between the entropy and $A$ has not yet been
reproduced. 

\subsubsection{Loop quantum gravity}

In loop quantum gravity, variables are used 
that are conceptually closer to Yang--Mills type of
variables. 
The loop variables have grown out of `Ashtekar's new variables'
(Ashtekar 1986), which
are defined as follows.
The role of the momentum variable 
is played by the densitized triad (dreibein)
\be
\lb{4.79}
E^a_i(x):=\sqrt{h}(x)e^a_i(x),
\ee
while the configuration variable is the connection
\be
\lb{4.93}
GA^i_a(x)=\Gamma^i_a(x)+\beta K^i_a(x).
\ee
Here, $a$ ($i$) denotes a space index (internal index);
$\Gamma^i_a(x)$ is the spin connection, and $K^i_a(x)$ is related to
the second fundamental form. The parameter $\beta$ is called
Barbero--Immirzi parameter and can assume any non-vanishing real
value. In loop quantum gravity, it is a free parameter. It may be
fixed by the requirement that the black-hole entropy calculated from
loop quantum cosmology coincides with the Bekenstein--Hawking
expression \eqref{SBH}. One thereby finds (from numerically solving an
equation) the
value $\beta=0.23753\ldots$, see Agullo {\em et al.} (2010) and the
references therein. It is in this context of interest to note that the
proportionality between entropy and area can only be obtained if the
microscopic degrees of freedom (the spin networks) are distinguishable 
(cf. Kiefer and Kolland~2008).
Classically, these variables obey the Poisson-bracket relation
\be
\lb{4.94}
\{A^i_a(x),E^b_j(y)\}=8\pi\beta\delta^i_j\delta^b_a\delta(x,y).
\ee
The loop variables are constructed from these variables in a non-local
fashion. The new connection variable is the holonomy $U[A,\alpha]$,
which is a path-ordered exponential of the integral over the
connection \eqref{4.93} around a loop $\alpha$. In the quantum theory,
it acts on wave functionals as
\be
\lb{6.20}
\hat{U}[A,\alpha]\Psi_S[A]= U[A,\alpha]\Psi_S[A].
\ee
The new momentum variable is the flux of the densitized triad through
a two-dimensional surface ${\mathcal S}$ 
bounded by the loop. Its operator version reads
\be
\lb{6.21}
\hat{E}_i[{\mathcal S}]:= -8\pi\beta\hbar\I\int_{\mathcal S}\D\sigma^1
\D\sigma^2\ n_a(\vec{\sigma})\frac{\delta}{\delta A^i_a[\X(\vec{\sigma})]},
\ee
where the embedding of the surface is given by
$(\sigma^1,\sigma^2)\equiv\vec{\sigma} 
\mapsto x^a(\sigma^1,\sigma^2)$. 
The variables obey the commutation relations
\bdm
\left[\hat{U}[A,\alpha],\hat{E}_i[{\mathcal S}]\right]
=\I l_{\rm P}^2\beta\iota(\alpha,{\mathcal S})U[\alpha_1,A]\tau_i
U[\alpha_2,A],
\edm
where $\iota(\alpha,{\mathcal S})=\pm 1, 0$ is the `intersection number',
which depends on the orientation of $\alpha$ and ${\mathcal S}$.
Given certain mild assumptions, the holonomy-flux representation is
unique and gives rise to a unique Hilbert-space structure at the
kinematical level, that is, before the constraints are imposed. 

At this level, one can define an area operator, for which one obtains
the following spectrum:
\be
\lb{6.31}
\hat{\mathcal A}[{\mathcal S}]\Psi_S[A]=8\pi\beta l_{\rm P}^2
\sum_{{\rm P}\in S\cap{\mathcal S}}\sqrt{j_{\rm P}(j_{\rm P}+1)}\Psi_S[A]
=: A[{\mathcal S}]\Psi_S[A].
\ee
Here, P denotes the intersection points between the spin-network S and
the surface ${\mathcal S}$, and the $j_{\rm P}$ can assume integer and
half-integer values (arising from the use of the group SU(2) for the
triads). There thus exists a minimal `quantum of action' of the order
of $\beta$ times the Planck-length squared. 
Comprehensive discussions of loop quantum gravity can be found in 
Gambini and Pullin (2011), Rovelli (2004), Ashtekar and Lewandowski
(2004), and Thiemann (2007). 

Here, we have been mainly concerned with the canonical version of loop
quantum gravity. But there
exists also a covariant version: it
corresponds to a path-integral formulation, through which the spin
networks are evolved `in time'. It is called the 
spin-foam approach, cf. Rovelli (2013) and the references therein.

\subsection{String theory}

String theory is fundamentally different from the approaches discussed
so far. It is not a direct quantization of GR or any other classical theory of
gravity. It is an example for a unified quantum theory of all
interactions. Gravity, as well as the other
known interactions, only emerges in an appropriate limit. 
(This is why string theory is an example of `emergent gravity'.) Strings are
one-dimensional objects characterized by a dimensionful parameter
$\alpha'$ or the string length $l_{\rm s}=\sqrt{2\alpha'\hbar}$.
In spacetime, it forms a two-dimensional surface,
the worldsheet. Closer inspection of the theory exhibits also the
presence of higher-dimensional objects called D-branes, which are as
important as the strings themselves, cf. Blumenhagen {\em et al.} (2013). 

String theory necessarily contains gravity, because the graviton
appears as an excitation of closed strings. It is through this
appearance that a connection to covariant quantum gravity discussed
above can be made. String theory also includes gauge theories, since
the corresponding gauge bosons are found in the spectrum. It also
requires the presence of supersymmetry for a consistent
formulation. Fermions are thus an important ingredient of string
theory. One recognizes that gravity, other fields, and matter appear
on the same footing.   

Because of reparametrization invariance on the worldsheet, string
theory also possesses constraint equations. The constraints do,
however, not close, but contain a central term on the right-hand
side. This corresponds to the presence of an anomaly (connected with
Weyl transformations). The vanishing of this anomaly can be achieved
if ghost fields are added that gain a central term which cancels the
original one. The important point is that this works only in a
particular number $D$ of dimensions: $D=26$ for the bosonic string,
and $D=10$ for the superstring (or $D=11$ in M-theory). The presence
of higher spacetime 
dimensions is an essential ingredient of string theory.   

Let us consider, for simplicity, the bosonic string. Its quantization
is usually performed through the Euclidean path integral
\be
\lb{9.2.2}
Z=\int{\mathcal D}X{\mathcal D}h\ \E^{-S_{\rm P}},
\ee
where $X$ and $h$ are a shorthand for the embedding variables and the
worldsheet metric, respectively. The action in the exponent is the
`Polyakov action', which is an action defined on the
worldsheet. Besides the dynamical variables $X$ and $h$, it contains
various background fields on spacetime, among them the metric of the
embedding space and a scalar field called dilaton. It is obvious that
this formulation is not background independent. In as much string
theory (or M-theory) can be formulated in a fully background
independent way, is a controversial issue. It has been argued that
partial background independence is implemented in the context of the
AdS/CFT conjecture, see Sec.~4 below.

If the string propagates in a curved spacetime with metric
$g_{\mu\nu}$, the demand for the absence of a Weyl anomaly leads to
consistency equations that correspond (up to terms of order $\alpha'$)
to the Einstein equations for the background fields. These equations
can be obtained from an effective action of the form
\bea
\lb{9.2.14}
S_{\rm eff} & \propto & 
\int\D^Dx\ \sqrt{-g}\ \E^{-2\Phi}\Big(R-\frac{2(D-26)}
{3\alpha'}-\frac{1}{12}H_{\mu\nu\rho}H^{\mu\nu\rho}
\nonumber\\
& & \; \;\;\;\; +4\nabla_{\mu}\Phi\nabla^{\mu}\Phi+{\mathcal O}(\alpha') 
\Big) ,
\eea
where $\Phi$ is the dilaton, $R$ the Ricci scalar corresponding to 
$g_{\mu\nu}$, and $H_{\mu\nu\rho}$ the field strength associated with
an antisymmetric tensor field (which in $D=4$ would be the axion). 
This is the second connection of string theory with gravity, after the
appearance of the graviton as a string excitation.
A recent comprehensive overview of string theory is Blumenhagen {\em
  et al.} (2013).

Attempts were made in string theory to derive the Bekenstein--Hawking
entropy from a microscopic counting of states. In this context, the
D-branes turned out to be of central importance. Counting D-brane
states for extremal and close-to-extremal string black holes, one was
able to derive \eqref{SBH} including the precise prefactor
(Strominger and Vafa~1996, Horowitz 1998). For the Schwarzschild black
hole, however, such a derivation is elusive.

Originally, string theory was devised as a `theory of everything' in
the strict sense. This means that the hope was entertained to derive
all known physical laws including all parameters and coupling
constants from this fundamental theory. So far, this hope remains
unfulfilled. Moreover, there are indications that this may not be
possible at all, due to the `landscape problem': string theory
seems to lead to many ground states at the effective level, the number
exceeding $10^{500}$ (Douglas~2003). Without any further idea, it
seems that a selection can be made only on the basis of the anthropic
principle, cf. Carr (2007). If this were the case, string theory would no longer
be a predictive theory in the traditional sense.

\subsection{Other approaches}

Besides the approaches mentioned so far, there exist further
approaches which are either meant to be separate quantum theories of
the gravitational field or candidates for a fundamental quantum theory
of all interactions. Some of them have grown out of one of the above
approaches, others have been devised from scratch. It is interesting
to note that most of these alternatives start from {\em discrete
  structures} at the microscopic level. Among them are such approaches
as causal sets, spin foams, group field theory, quantum topology, and
theories invoking some type of non-commutative geometry. Most of them
have not been developed as far as the approaches discussed above. An
overview can be found in Oriti (2009).

\section{Quantum cosmology}

Quantum cosmology is the application of quantum theory to the universe
as a whole. Conceptually, this corresponds to the problem of
formulating a quantum theory for a closed system from within, without
reference to any external observers or measurement agencies. In its
concrete formulation, it demands for a quantum theory of gravity,
since gravity is the dominating interaction at large scales. On the
one hand, quantum cosmology may serve as a testbed for quantum gravity
in a mathematically simpler setting. This concerns, in particular, the
conceptual questions which are of concern here. On the other hand,
quantum cosmology may be directly relevant for an understanding of the
real Universe. General introductions into quantum cosmology include
Coule (2005), Halliwell (1991), Kiefer (2012), Kiefer and Sandh\"ofer
(2008) and Wiltshire (1996). A
discussion in the general context of cosmology can be found in Montani
{\em et al.} (2011). Supersymmetric quantum cosmology is discussed at
depth in Moniz (2010). An introduction to loop quantum cosmology is
Bojowald (2011). A comparison of standard quantum cosmology with loop
quantum cosmology can be found in Bojowald {\em et al.} (2010). 

Quantum cosmology is usually discussed for homogeneous models
(the models are then called minisuperspace models). The
simplest case is to assume also isotropy. Then,
the line element for the classical spacetime metric is given by
\be
ds^2=-N(t)^2dt^2+a(t)^2d\Omega_3^2\ ,
\ee
where $d\Omega_3^2$ is the line-element of an constant curvature space 
with curvature index $k=0,\pm 1$. In order to consider a matter degree
of freedom, a homogeneous scalar field $\phi$ with potential $V(\phi)$
is added. 

In this setting, the momentum
constraints \eqref{5.22} are identically fulfilled. The
Wheeler--DeWitt equation \eqref{5.21} becomes a two-dimensional
partial differential equation for a wave function $\psi(a,\phi)$,
\be
\lb{wdw1}
\left(\frac{\hbar^2\kappa^2}{12}a\frac{\partial}{\partial a}
a\frac{\partial}{\partial a}-\frac{\hbar^2}{2}\frac{\partial^2}
{\partial\phi^2}+a^6\left(V(\phi)
+\frac{\Lambda}{\kappa^2}\right)-\frac{3ka^4}{\kappa^2}\right)
\Psi(a,\phi)=0 ,
\ee
where $\Lambda$ is the cosmological constant, and $\kappa^2=8\pi G$. 
Introducing $\alpha\equiv\ln a$
(which has the advantage to have a range from $-\infty$ to $+\infty$),
one obtains the following equation 
\be
\lb{wdw2}
\left(\frac{\hbar^2\kappa^2}{12}\frac{\partial^2}{\partial\alpha^2}-
\frac{\hbar^2}{2}\frac{\partial^2}{\partial\phi^2}+ 
e^{6\alpha}\left(V\left(\phi\right)+
\frac{\Lambda}{\kappa^2}\right)
-3e^{4\alpha}\frac{k}{\kappa^2}\right)\Psi(\alpha,\phi)=0 
\ . 
\ee
Most versions of quantum cosmology take Einstein's theory as the
classical starting point. More general approaches include
supersymmetric quantum cosmology (Moniz 2010), string quantum
cosmology (2003), non-commutative
quantum cosmology (see e.g. Bastos {\em et al.} 2008), 
Ho\v{r}ava--Lifshitz quantum cosmology (Bertolami and Zarro 2011),
and third-quantized cosmology (Kim 2013).

In loop quantum cosmology, features from full loop quantum gravity are
imposed on cosmological models (Bojowald 2011).
Since one of the main features is the discrete nature of geometric
operators, the Wheeler--DeWitt equation \eqref{wdw1} is replaced by a
difference equation. This difference equation becomes
indistinguishable from the Wheeler--DeWitt equation at scales
exceeding the Planck length, at least in certain models. Because the
difference equation is difficult to solve in general, one makes
heavily use of an effective theory (Bojowald 2012).

Many features of quantum cosmology are discussed in the limit when
the solution of the Wheeler--DeWitt equation assumes a semiclassical
or WKB form (Halliwell 1991). This holds, in particular, when the
no-boundary proposal or the tunnelling proposal is investigated for
concrete models, see below. It has even been suggested that the wave function of
the universe be interpreted only in the WKB limit, because only then a
time parameter and an approximate (functional) Schr\"odinger equation
is available (Vilenkin 1989). 

This can lead to a conceptual confusion. Implications for the meaning of
the quantum cosmological wave functions should be derived as much as
possible from exact solutions. This is because the WKB approximation
breaks down in many interesting situations, even for a universe of
macroscopic size. One example is a closed Friedmann universe with a massive
scalar field (Kiefer 1988). The reason is the following. For a
classically recollapsing universe, one must impose the boundary
condition that the wave function go to zero for large scale factors,
$\Psi\to0$ for large $a$. As a consequence, narrow wave packets do not
remain narrow because of the ensuing scattering phase shifts of the
partial waves (that occur in the expansion of the wave function 
into basis states) from the turning point. The correspondence to the
classical model can only be understood if the quantum-to-classical
transition in the sense of decoherence (see below) is invoked.  

Another example is the case of classically chaotic cosmologies, see,
for example, Calzetta and Gonzalez (1995) and Cornish and Shellard
(1998). Here, one can see that the WKB approximation breaks down in
many situations. This is, of course, a situation well known from
quantum mechanics. One of the moons of the planet Saturn, Hyperion,
exhibits chaotic rotational motion. Treating it quantum mechanically,
one recognizes that the semiclassical approximation breaks down and
that Hyperion is expected to be in an extremely nonclassical state of
rotation (in contrast to what is observed). This apparent conflict
between theory and observation can be understood by invoking the
influence of additional degrees of freedom in the sense of
decoherence, see Zurek and Paz (1995) and Sec.~3.3.4.3 in Joos {\em et
  al}. (2003). The same mechanism should cure the situation
for classically chaotic cosmologies (Calzetta 2012).

\section{The problem of time}

The problem of time is one of the major conceptual issues in the
search for a quantum theory of gravity (Anderson~2012, Isham~1993,
Kucha\v{r}~1992). 
As was already emphasized above, time is treated
differently in quantum theory and in general relativity. Whereas it is
absolute in the first case, it is dynamical in the second. This is
the reason why GR is background independent -- a major feature to
consider in the quantization of gravity.

Background independence is only part of the problem of time. If one  
applies the standard quantization rules to GR, spacetime disappears
and only space remains. This can be understood in simple
terms. Classically, spacetime corresponds to what is a particle
trajectory in mechanics. Upon quantization, the trajectory vanishes and only the
position remains. In GR, spacetime vanishes and only the three-metric
remains. Time has disappeared. 

This is explicitly seen by the timeless nature of the Wheeler--DeWitt
equation \eqref{5.21} or its analogue in loop quantum gravity. How can
one interpret such a situation? Equation \eqref{5.21} results from a
classical constraint in which all momenta occur quadratically. If one
could reformulate this constraint in a way where one canonical momentum
appears linearly, its quantized form would exhibit a
Schr\"odinger-type equation. Concretely, one would have classically
\be
\lb{SolvedHamiltonian}
\mathcal{P}_A+h_A=0\ ,
\ee
where $\mathcal{P}_A$ is a momentum for which
one can solve the constraint; $h_A$ simply stands for the remaining terms.
Upon quantization one obtains
\be
\lb{effectiveS}
\mathrm{i}\hbar\frac{\delta\Psi}{\delta q_A}=\widehat{h}_A\Psi.
\ee
This is of a Schr\"odinger form. 
The dynamics of this equation is
  in general inequivalent to the Wheeler--DeWitt equation.
Usually, $h_A$ is referred to as a `physical Hamiltonian' because it
actually describes an evolution in a {\it physical} parameter, namely the
coordinate $q_A$ conjugate to $\mathcal{P}_A$. 

There are many problems with such a `choice of time before
quantization' (Kucha\v{r}~1992). The constraints of GR cannot be put
globally into the form \eqref{SolvedHamiltonian} (Torre 1993). In most
cases, the operator $\widehat{h}_A$ cannot be defined rigorously. 
It has therefore been suggested to introduce dust matter with the sole
purpose to define a standard of time (Brown and Kucha\v{r}~1995). More
recently, a massless scalar field is used in loop quantum cosmology 
for this purpose, cf. Ashtekar and Singh (2011). One can also define
such an internal time from an electric field (Alexander {\em et
  al.}~2012). At an effective level, one can choose different local internal
times within the same model (Bojowald {\em et al.}~2011). It is,
however, not very satisfactory to adopt the concept of time to the
particular model under consideration.

Most of these discussions make use of the Schr\"odinger picture of
quantum theory. An interesting perspective on the problem of time
using the Heisenberg picture is presented in Rovelli (1991). It leads
to the notion of an `evolving constant of motion', which unifies the
intuitive idea of an evolution with the timelessness of quantum
gravity; the evolution proceeds with respect to a physical `clock'
variable. An explicit construction of evolving constants in a concrete
non-trivial model can be found, for example, in Montesinos {\em et
  al.} (1999).

A recent approach to treat the problem of time is shape dynamics, see
Barbour {\em et al.} (2013) and the references therein. In this
approach, three-dimensional conformal invariance plays the central
role. The configuration space is conformal superspace (the
geometrodynamic shape space) times ${\mathbb R}^+$, the second part
coming from the volume of three-space. Spacetime foliation invariance
at the classical level has been lost. In shape dynamics, the variables
are naturally separated into dimensionless true degrees of freedom and
a single variable that serves the role of time. This approach is
motivated by a similar approach in particle mechanics (Barbour and
Bertotti~1982), see also Barbour (2000) and Anderson (2011). The
consequences of this for the quantum version of 
shape dynamics have still to be explored. 

Independent of these investigations addressing the concept of time
in full quantum gravity, it is obvious that the limit of quantum field
theory in curved spacetime (in which time as part of spacetime exists)
must be recovered in an appropriate limit. How this is achieved, is
not clear in all of the above approaches. It is most transparent for
the Wheeler--DeWitt equation, as I will be briefly explaining now.

In the semiclassical approximation to the Wheeler--DeWitt equation,
one starts with the ansatz
\be
\lb{ansatz}
        \vert\Psi[h_{ab}]\rangle=C[h_{ab}]
        \E^{\I m_{\rm P}^2S[h_{ab}]}|\psi [h_{ab}]\rangle
\ee
and performs an expansion with respect to the inverse Planck mass squared
$m_{\rm P}^{-2}$. This is inserted into (\ref{5.21}) and \eqref{5.22},
and consecutive 
orders in this expansion are considered, see Kiefer and Singh (1991),
Bertoni {\em et al.} (1996),
and Barvinsky and Kiefer (1998). This is 
close to the Born--Oppenheimer (BO) approximation scheme in
molecular physics. 
In (\ref{ansatz}), $h_{ab}$ denotes again the three-metric, and the Dirac
bra-ket notation refers to the non-gravitational fields, for which the
usual Hilbert-space structure is assumed.

The highest orders of the BO scheme lead to the following picture.
One evaluates the `matter wave function' $|\psi[h_{ab}]\rangle$ along a
solution of the classical Einstein equations, $h_{ab}({\mathbf x},t)$,
which corresponds to a chosen solution $S[h_{ab}]$ of the Hamilton--Jacobi
equations.
One can then define 
\bdm
        \dot{h}_{ab}=NG_{abcd}
        \frac{\delta S}{\delta h_{cd}}+
        2D_{(a}{N_{b)}},
\edm
where $N$ is the lapse function and $N^a$ is the shift vector; their
choice reflects the chosen foliation and coordinate assignment. 
In this way, one can recover a classical spacetime as an
approximation, a spacetime that satisfies Einstein's equations in this
limit.  
The time derivative of the matter wave function is then defined by 
\bdm
        \frac{\partial}{\partial t}\,|\psi(t)\rangle:=
        \int \D^3 x \,\dot{h}_{ab}({\bf x},t)\,
        \frac{\delta}{\delta h_{ab}({\bf x})}
        |\psi[h_{ab}]\rangle, 
\edm
where the notation $|\psi(t)\rangle$ means $|\psi [h_{ab}]\rangle$
evaluated along the chosen spacetime with the chosen foliation.
This leads to a functional Schr\"odinger equation for quantized
matter fields in the chosen external classical gravitational field,
        \begin{eqnarray}
\lb{Schrodinger}
        \I\hbar\frac{\partial}{\partial t}\,
        |\psi(t)\rangle &=& \hat{H}{}^{\rm m}|\psi(t)\rangle,\\
        \hat{H}{}^{\rm m} &:=&
        \int \D^3 x \left\{ N({\bf x})
        \hat{\mathcal H}{}^{\rm m}_{\perp}({\bf x})+
        N^a({\bf x})\hat{\mathcal H}{}^{\rm m}_a({\bf x})\right\},
               \end{eqnarray}
where $\hat{H}{}^{\rm m}$ denotes the 
 matter-field Hamiltonian in the Schr\"odinger
picture, which depends parametrically on the (generally non-static) metric
coefficients of the curved space--time background recovered from
$S[h_{ab}]$. The `WKB time' $t$ controls the dynamics in this approximation.

In the semiclassical limit, one thus finds a Schr\"odinger equation of
the form \eqref{effectiveS}, without using an artifical scalar field
or dust matter that are often introduced with the sole purpose of
defining a time at the exact level, cf. \eqref{effectiveS}. From an
empirical point of view, nothing more is 
needed, because we do not have any access so far to a regime where the
semiclassical approximation is not valid. In this limit, standard
quantum theory with all its machinery (Hilbert space, probability
interpretation) emerges, and it is totally unclear whether this
machinery is needed beyond the semiclassical approximation.

Proceeding to the next order of the $m_{\rm P}^{-2}$-expansion, one
can derive quantum gravitational corrections to the functional
Schr\"odinger equation \eqref{Schrodinger}. These may, in principle, lead to
observable contributions to the anisotropy spectrum of the cosmic
microwave background (CMB) anisotropy spectrum (Kiefer and Kr\"amer
2012, Bini {\em et al.} 2013), although they are at present too tiny. 

A major conceptual issue concerns the arrow of time
(Zeh~2007). Although our fundamental laws, as known so far, are
time-reversal invariant (or a slight generalization thereof), there
are classes of phenomena that exhibit a definite temporal
direction. This is expressed by the Second Law of thermodynamics. 
Is there a hope that the origin of this irreversibility can be found in
quantum gravity?

Before addressing this possibility, let us estimate how special our
Universe really is, that is, how large its entropy is compared with its
maximal possible entropy.
Roger Penrose has pointed out that the maximal
entropy for the observable Universe would be obtained if all its
matter were assembled into one black hole (Penrose 1981).
Taking the most recent
observational data, this gives the entropy (Kiefer~2009)
\be
\lb{Smaxnew}
S_{\rm max}\approx 1.8\times 10^{121} \ . 
\ee
(Here and below, $k_{\rm B}=1$.)
This may not yet be the maximal possible entropy. Our Universe
exhibits currently an acceleration caused by a
cosmological constant~$\Lambda$ or a dynamical dark energy. If it were
caused by ~$\Lambda$, it would expand 
forever, and the entropy in the far future would be dominated by the
entropy of the cosmological event horizon. This entropy is called
the `Gibbons--Hawking entropy' (Gibbons and Hawking 1977)
and leads to (Kiefer~2009)
\be
\lb{Sgh}
S_{\rm GH}=\frac{3\pi}{\Lambda l_{\rm P}^2}\approx 2.9\times
10^{122}\ ,
\ee
which is about one order of magnitude higher than (\ref{Smaxnew}).

Following the arguments in Penrose (1981), the `probability' for
our Universe can then be estimated as
\be
\frac{\exp(S)}{\exp(S_{\rm max})}\approx \frac{\exp(3.1\times
  10^{104})}{\exp(2.9\times 10^{122})}\approx \exp(-2.9\times
10^{122})\ .
\ee
Our Universe is thus very special indeed. It is much more special than
what would be estimated from the anthropic principle. 

Turning to quantum gravity, the question arises how one can derive an
arrow of time from a framework that is fundamentally timeless. A final
answer is not yet available, but various ideas exist (Zeh~2007,
Kiefer~2009). The Wheeler--DeWitt equation \eqref{5.21} does not
contain any external time parameter, but one can define an intrinsic
time from the hyperbolic
nature of this equation, which is entirely constructed from the
three-metric (Zeh~1988). In quantum cosmology, this intrinsic time is
the scale factor, cf. \eqref{wdw1} and \eqref{wdw2}. 
If one adds small inhomogeneous degrees of freedom to a Friedmann
model (Halliwell and Hawking 1985), the Wheeler--DeWitt equations
turns out to be of the form
\begin{equation}
\lb{3WdW2}
 \hat{H} \, \Psi = \left(\frac{2\pi G\hbar^2}{3}
\frac{\partial^2}{\partial\alpha^2} + \sum_i \, \left[
-\frac{\hbar^2}{2}
\frac{\partial^2}{\partial x_i^2}+\underbrace{V_i(\alpha,x_i)}_{\to 0\ 
{\rm for}\ \alpha
\rightarrow -\infty}\right]\right) \, \Psi = 0 \ ,
\end{equation}
where the $\{ x_i\}$ denote the inhomogeneous
degrees of freedom as well as neglected homogeneous ones;
$V_i(\alpha,x_i)$ are the corresponding potentials. One
recognizes immediately that this Wheeler--DeWitt equation is
hyperbolic with respect to the intrinsic time $\alpha$. Initial
conditions are thus most naturally formulated with respect to constant
$\alpha$. 

The important property of \eqref{3WdW2} is
that the potential becomes small for $\alpha\to -\infty$ (where the
classical singularities would occur), but complicated for increasing
$\alpha$. In the general case (not restricting to small
inhomogeneities), this may be further motivated by the
BKL-conjecture according to which spatial gradients become small
near a spacelike singularity (Belinskii {\em et al.} 1982). 
The Wheeler--DeWitt
equation thus possesses an asymmetry with respect to `intrinsic
time' $\alpha$. 
One can in particular impose the simple boundary
condition (Zeh~2007)
\begin{equation}
\lb{initialcondition}
\Psi \quad \stackrel{\alpha \, \to \, -\infty}{\longrightarrow}\
\psi_0(\alpha)\prod_i \psi_i(x_i)\ ,
\end{equation}
which means that the degrees of freedom are initially {\em not}
entangled. They will become entangled for increasing $\alpha$ because
then the coupling in the potential between $\alpha$ and the $\{ x_i\}$
becomes important. This leads to a positive entanglement entropy. In
the semiclassical limit, where a WKB time $t$ can be defined, there is
a correlation between $t$ and $\alpha$ which can explain the standard
Second Law, at least in principle. In this scenario, entropy increase
is correlated with scale-factor increase. Therefore, the arrow of time
would formally reverse at the turning point of a classically
recollapsing universe, although in a quantum scenario the classical
evolution would end there and no transition to a recollapsing phase
could ever be observed (Kiefer and Zeh~1995). Whether a boundary
condition of the form \eqref{initialcondition} follows as a necessary
requirement from the mathematical structure of the full theory or
not is an open issue.
Alternative ideas to the recovery of the arrow of time can be found in
Penrose (2009), Vilenkin (2013{\em a}), and the references therein. 
 
Our discussion about the problem of time was performed in the
framework of quantum 
geometrodynamics. Its principle features should also be applicable,
with some modifications, to loop quantum gravity. But what about
string theory?

String theory contains general relativity. As we have seen above, the
quantum equation that gives back Einstein's equation in the
semiclassical limit is the Wheeler--DeWitt equation \eqref{5.21}. One
would thus expect that the  Wheeler--DeWitt equation can be recovered
from string theory in the limit where the string constant $\alpha'$ is
small. Unfortunately, this has not been shown so far in any explicit manner.

It is clear, however, that the problem of time is the same in string
theory. New insights may be obtained, in addition, for the concept of
space. This 
is most clearly seen in the context of the AdS/CFT correspondence,
see, for example, Maldacena (2011) for a review. 
In short words, this correspondence states that 
non-perturbative string theory in
  a background spacetime which is asymptotically anti-de Sitter (AdS)
  is dual to a conformal field theory (CFT) defined in a flat
  spacetime of one dimension fewer (the boundary of the background
  spacetime). What really corresponds here are certain matrix elements
  and symmetries in the two theories; an equivalence at the level of
  the quantum states is not shown.
This correspondence can be considered as an intermediate step towards
a background-independent formulation of 
string theory, because the background metric enters only through boundary
conditions at infinity, cf. Blau and Theisen (2009). 

The AdS/CFT correspondence can also be interpreted as a 
realization of the `holographic principle', which states that the
information of a gravitating system is located on the boundary of this
system; the most prominent example is the expression \eqref{SBH} for
the entropy, which is given by the surface of the event horizon. In a
particular case, 
laws including gravity in $d=3$ are equivalent to laws
excluding gravity in $d=2$.
In a loose sense, space has then vanished, too (Maldacena~2005). 

Our discussion of the problem of time presented in this section is far
from complete. There exist, for example, interesting suggestions of a
thermodynamical origin of time in a gravitational context (Rovelli
1993). It has been shown that a statistical mechanics of generally
covariant quantum theories can be developed without a preferred time
and thus without a preferred notion of temperature (Montesinos and
Rovelli 2001).  


\section{Singularity avoidance and boundary conditions}

As we have mentioned at the beginning, classical GR predicts the
occurrence of singularities (Hawking and Penrose~1996). What can the
above approaches say about their fate in the quantum theory? In order
to discuss this, one first has to agree about a definition of
singularity avoidance in quantum gravity. Since such an agreement does
not yet exist, one has to study heuristic expectations
Already DeWitt (1967) has speculated that a classical singularity is avoided
if the wave function vanishes at the corresponding region in
configuration space,
\be
\lb{8.3.0}
\Psi\left[{}^{(3)} {\mathcal G}_{\rm sing}\right]=0.
\ee
One example where this condition can be implemented concerns the fate
of a singularity called `big brake' (Kamenshchik {\em et al.}~2007).
This occurs for an
equation of state of the form $p=A/\rho$, $A>0$, called
`anti-Chaplygin gas'. 
For a Friedmann
universe with scale factor $a(t)$ and a scalar field $\phi(t)$, this
equation of state can be realized by the potential
\bdm
V(\phi)=V_0\left(\sinh{\left(\sqrt{3\kappa^2}|\phi|\right)}-
\frac1{\sinh{\left(\sqrt{3\kappa^2}|\phi|\right)}}\right)\ ; \ V_0=\sqrt{A/4}\ ,
\edm
where $\kappa^2=8\pi G$.
The classical dynamics 
develops a pressure singularity (only $\ddot{a}(t)$ becomes singular)
and comes to an abrupt halt in the future (`big brake').

The Wheeler--DeWitt equation for this model reads
\bea
& & \frac{\hbar^2}{2}\left(\frac{\kappa^2}{6}\frac{\partial^2}
{\partial\alpha^2}-\frac{\partial^2}{\partial\phi^2}\right)\Psi\left(\alpha,
  \phi\right)\nonumber\\ & & \;
+V_0e^{6\alpha}\left(\sinh{\left(\sqrt{3\kappa^2}|\phi|\right)}
-\frac1{\sinh{\left(\sqrt{3\kappa^2}|\phi|\right)}}\right)\Psi\left(\alpha, 
  \phi\right)=0\ ,
\eea
where $\alpha=\ln a$, and Laplace--Beltrami factor ordering
has been used. 
The vicinity of the big-brake singularity is the region of small
$\phi$; we can therefore use the approximation
\bdm
\frac{\hbar^2}2\left(\frac{\kappa^2}{6}\frac{\partial^2}{\partial\alpha^2}
-\frac{\partial^2}{\partial\phi^2}\right)\Psi\left(\alpha,
  \phi\right)-\frac{\tilde{V_0}}{|\phi|}e^{6\alpha}\Psi\left(\alpha,
  \phi\right)=0\ ,
\edm
where $\tilde{V_0}={V_0}/{3\kappa^2}$.

It was shown in Kamenshchik {\em et al.} (2007) that all normalizable
solutions are of the form 
\bdm
\Psi\left(\alpha,\phi\right)=
\sum_{k=1}^{\infty}A(k)k^{-3/2}\mathrm{K}_0
\left(\frac{1}{\sqrt{6}}\frac{V_{\alpha}}{\hbar^2k\kappa}\right)
\times\left(2\frac{V_\alpha}{k}|\phi|\right)
e^{-\frac{V_\alpha}{k|\phi|}}\mathrm{L}^1_{k-1}
\left(2\frac{V_\alpha}{k}|\phi|\right)\ ,
\edm
where $\mathrm{K}_0$ is a Bessel function, $\mathrm{L}^1_{k-1}$ are Laguerre
polynoms, and $V_\alpha\equiv\tilde{V_0}e^{6\alpha}$. They all {\em
  vanish} at the classical singularity. This model therefore
implements DeWitt's criterium above. The same holds in more general
situations of this type (Bouhmadi-L\'opez {\em et al.} 2009). In a
somewhat different approach, the big brake singularity is not avoided
(Kamenshchik and Manti~2012). Singularity avoidance in the case of a
`big rip', which can occur for phantom fields, is discussed in
D\c{a}browski {\em et al.} (2006). 

The vanishing of the wave function at the classical singularity plays
also a role in the treatment of
supersymmetric quantum cosmological billiards (Kleinschmidt {\em et
  al.}~2009). In $D=11$ supergravity, one can employ near
a spacelike singularity such a description based on the
Kac--Moody group $E_{10}$ and derive the corresponding
Wheeler--DeWitt equation. It was found there, too, that 
$\Psi\to 0$ near the singularity and that DeWitt's criterium is
fulfilled. Singularity avoidance for the Wheeler--DeWitt equation is
also achieved when the Bohm interpretation is used (Pinto-Neto {\em et
  al.}~2012). Singularity avoidance was also discussed using `wavelet
quantization' instead of canonical quantization (Bergeron {\em et al.}
2013); there the occurrence of a repulsive potential was found. 
 
Singularity avoidance also occurs in the framework of
loop quantum cosmology, although in a somewhat different way
(Bojowald~2011). The big-bang singularity can be avoided by solutions
of the difference equation that replaces the Wheeler--DeWitt
equation. The avoidance can also be achieved by the occurrence of a
bounce in the effective Friedmann equations. These results strongly
indicate that singularities are avoided in loop quantum
cosmology, although no general theorems exist (Ashtekar and
Singh~2011). Possible singularity avoidance can also be discussed for
black-hole singularities and for naked singularities (Joshi~2013). 

The question of singularity avoidance is closely connected with the
role of boundary conditions in quantum cosmology. Let us thus here
have a brief look at the conceptual side of this issue.

The Wheeler--DeWitt equation \eqref{wdw2} is of hyperbolic nature with
respect to $\alpha\equiv \ln a$. It makes thus sense to specify the wave
function and its derivative at constant $\alpha$. In the case of an
open universe, this is fine. In the case of a closed universe,
however, one has to impose the boundary condition that the wave
function vanishes at $\alpha\to\infty$. The existence of a
solution in this case may then lead to a restriction on the allowed
values for the parameters of the model, such as the cosmological
constant or the mass of a scalar field, or may allow no solution at all. 

Another type of boundary conditions makes use of the path-integral
formulation \eqref{pathgrav}. In 1982, Hawking formulated his
`no-boundary condition' for the wave function (Hawking 1982), which
was then elaborated in Hartle and Hawking (1983) and Hawking (1984). 
The wave function is given by the Euclidean path integral
\be
\lb{8.3.1}
\Psi[h_{ab},\Phi,\Sigma]=\sum_{\mathcal M}\nu({\mathcal M})
\int_{\mathcal M}{\mathcal D}g{\mathcal D}\Phi\
\E^{-S_{\rm E}[g_{\mu\nu},\Phi]}\ .
\ee
The sum over $ {\mathcal M}$ expresses the sum over all four-manifolds
with measure $\nu({\mathcal M})$ (which actually cannot be
performed). The no-boundary condition states 
that -- apart from the boundary where the three metric $h_{ab}$ is
specified -- there is no other boundary on which initial
conditions have to be specified. 

Originally, the hope was entertained that the no-boundary proposal
leads to a unique wave function (or a small class of wave functions)
and that the classical big-bang singularity is smoothed out by the
absence of the initial boundary. It was, however, later realized that
there are, in fact, many solutions, and that the integration has to be
performed over complex metrics (see e.g. Halliwell and Louko~1990 and
Kiefer~1991). Moreover, the path integral can usually only be
evaluated in a semiclassical limit (using the saddle-point
approximation), so it is hard to make a general statement about
singularity avoidance. 

In a Friedmann model with scale factor $a$ and a scalar field $\phi$
with a potential $V(\phi)$, the no-boundary condition gives the
semiclassical solution (Hawking~1984) 
\be
\lb{8.3.6}
\psi_{\rm NB}\propto \left(a^2V(\phi)-1\right)^{-1/4}
\exp\left(\frac{1}{3V(\phi)}
\right)\cos\left(\frac{(a^2V(\phi)-1)^{3/2}}{3V(\phi)}-\frac{\pi}{4}\right)\ .
\ee
We note that the no-boundary wave function is always real. The form
\eqref{8.3.6} corresponds to the superposition of an expanding and a
recollapsing universe.

Another prominent boundary condition is the tunnelling proposal
(see Vi\-len\-kin~2003 and the references therein). It was originally
defined by the choice of taking 
`outgoing' solutions at singular boundaries of superspace. The term
`outgoing' is somewhat misleading, because the sign of the imaginary
unit has no absolute meaning in the absence of external time
(Zeh~1988). What one can say is that a complex solution is chosen, in
contrast to the real solution found from the no-boundary proposal. In
the same model, one obtains for the tunnelling proposal instead of
\eqref{8.3.6} the expression 
\be
\lb{8.3.9}
\psi_{\rm T}\propto (a^2V(\phi)-1)^{-1/4}\exp\left(-\frac{1}{3V(\phi)}
\right)\exp\left(-\frac{\I}{3V(\phi)}(a^2V(\phi)-1)^{3/2}\right).
\ee
Considering the conserved Klein--Gordon type of current
\be
\lb{8.3.7}
j=\frac{\I}{2}(\psi^*\nabla\psi-\psi\nabla\psi^*) , \quad \nabla j=0,
\ee
where $\nabla$ denotes the derivatives in minisuperspace, one finds for
a WKB solution of the form $\psi\approx C\exp(\I S)$ the expression
\be
\lb{8.3.8}
j\approx -\vert C\vert^2\nabla S\ .
\ee
The tunnelling proposal states that this current should point outwards
at large $a$ and $\phi$ (provided, of course, that $\psi$ is of WKB form 
there). If $\psi$ were real (as is the case in the no-boundary proposal),
the current would vanish. Again, the wave function is of semiclassical
form. It has been suggested that it may only be interpreted in this
limit (Vilenkin 1989). Other boundary conditions include the
SIC-proposal put forward by Conradi and Zeh (1991).

An interesting application of these boundary conditions concerns the
prediction of an inflationary phase for the early universe. Here, it
seems that the tunnelling wave function favours such a phase, while
the no-boundary condition disfavours it (Barvinsky {\em et al.}
2010). In the context of the string theory landscape, the no-boundary
proposal was applied to inflation, even leading to a prediction
for the spectral index of the CMB spectrum (Hartle {\em et al.}~2011).   


\section{General interpretation of quantum theory}

Quantum cosmology can shed some light on the problem of interpreting
quantum theory in general. After all, a quantum universe possesses by
definition no external classical measuring agency. It is does not
possible to apply the Copenhagen interpretation, which presumes the
existence of classical realms from the outset. 

Since all the approaches discussed above preserve the linearity of the
formalism, the superposition principle remains valid, and with it the
measurement problem. In most investigations, the Everett
interpretation (Everett 1957) is applied to quantum cosmology. In this
interpretation, all components of the wave function are equally real.
This view is already reflected by the words of Bryce DeWitt in DeWitt
(1967):
\begin{quote}
Everett's view of the world is a very natural one to adopt in the
quantum theory of gravity, where one is accustomed to speak without
embarassment of the `wave function of the universe.' It is possible
that Everett's view is not only natural but essential.
\end{quote}
Alternative interpretations include the Bohm interpretation, see
Pinto-Neto {\em et al.} (2012) and the references therein.

Quantum cosmology (and quantum theory in general) can be consistently
interpreted if the Everett interpretation is used together with the
process of decoherence (Joos {\em et al.} 2003). Decoherence is the
irreversible emergence of classical properties from the unavoidable
interaction with the `environment' (meaning irrelevant or negligible
degrees of freedom in configuration space). In quantum cosmology, the
irrelevant degrees of freedom include tiny gravitational waves and
density fluctuations (Zeh 1986). Their interaction with the scale
factor and global matter fields transform them into variables that
behave classically, see, for example, Kiefer (1987) and Barvinsky {\em
et al.} (1999). An application of decoherence to the superposition of
triad superpositions in loop quantum cosmology can be found in Kiefer
and Schell (2013).

Once the `background variables' such as $a$ or $\phi$ have been
rendered classical by decoherence, the question arises what happens to
the primordial quantum fluctuations out of which -- according to the
inflationary scenario -- all structure in the Universe emerges. 
Here, again, decoherence plays the decisive role, see Kiefer and
Polarski (2008) and the references therein. The quantum fluctuations
assume classical behaviour by the interaction with small perturbations
that arises either from other fields or from a self-coupling interaction.
The decoherence time typically turns out to be of the order
\be
t_{\rm d}\sim \frac{H_{\rm I}}{g}\ ,
\ee
where $g$ is a dimensionless coupling constant of the interaction
with the other fields causing decoherence, and $H_{\rm I}$ is the
Hubble parameter of inflation. The ensuing
coarse-graining brought about by the decohering fields causes an
entropy increase for the primordial fluctuations (Kiefer {\em et al.}
2007). All of this is in accordance with current observations of the CMB
anisotropies. 

These considerations can, in principle, be extended to the concept
of the multiverse, as it arises, for example, from the string
landscape or from inflation (see e.g. Carr 2007 and
Vilenkin 2013{\em b}), but additional problems emerge 
(such as the measure problem) that have
so far not been satisfactorily been dealt with.

As we have mentioned in Sec.~1, the linearity of quantum theory may
break down if gravity becomes important. In this case, a
collapse of the wave function may occur, which destroys all the
components of the 
wave function except one, see, for example, Landau {\em et al.} (2012).
But as long as there is no empirical
hint for the breakdown of linearity, the above scenario provides a
consistent and minimalistic (in the sense of mathematical structure)
picture of the quantum-to-classical 
transition in quantum cosmology.  

In addition to the many formal and mathematical problems, conceptual
problems form 
a major obstacle for the final construction of a quantum theory of
gravity and its application to cosmology. They may, however, also
provide the key 
for the construction of such a theory. Whether or when this will
happen is, however, an open question. 




\thebibliography{0}

\bibitem{} Agullo, I., Barbero~G., J. F., Borja, E. F., Diaz-Polo, J.,
and Villase\~{n}nor, E. J. S. (2010). 
Detailed black hole state counting in loop quantum gravity.
{\em Phys. Rev. D}, {\bf 82}, 084029. 


\bibitem{} Albers, M., Kiefer, C., and Reginatto,
           M. (2008). Measurement analysis and quantum gravity. {\em
           Phys. Rev. D}, {\bf 78}, 064051.

\bibitem{} Alexander, S., Bojowald, M., Marciano, A., and Simpson,
  D. (2012). Electric time in quantum cosmology. arXiv:1212.2204v1
  [gr-qc].  

\bibitem{} Almheiri, A., Marolf, D., Polchinski, J., and Sully,
  J. (2013).  	
Black holes: complementarity or firewalls? {\em J. High Energy Phys.},
02 (2013) 062.

\bibitem{} Ambj\o rn, J., Goerlich, A., Jurkiewicz, J., and Loll,
  R. (2013). Quantum gravity via causal dynamical triangulations.
  arXiv:1302.2173v1 [hep-th].

\bibitem{} Amelino-Camelia, G. and Kowalski-Glikman (eds.) (2005).
           {\em Planck scale effects in astrophysics and cosmology}.
           Lecture Notes in Physics~669. Springer, Berlin.

\bibitem{} Anderson, E. (2011). The problem of time and quantum
  cosmology in the relational particle mechanics arena. 
  arXiv:1111.1472v3 [gr-qc].

\bibitem{} Anderson, E. (2012). Problem of time in quantum gravity. 
           {\em Ann. Phys. (Berlin)}, {\bf 524}, 757--86.

\bibitem{} Anderson, J. (1967). {\em Principles of relativity physics}.
           Academic Press, New York.

\bibitem{} Ashtekar, A. (1986). New variables for classical and quantum
           gravity. {\em Phys. Rev. Lett.}, {\bf 57}, 2244--7.

\bibitem{} Ashtekar, A. and Lewandowski, J. (2004).
           Background independent quantum gravity: a status report.
           {\em Class. Quantum Grav.}, {\bf 21}, R53--152.

\bibitem{} Ashtekar, A. and Singh, P. (2011).            	
Loop quantum cosmology: a status report.
{\em Class. Quantum Grav.}, {\bf 28}, 213001.

\bibitem{} Barbour, J. B. (2000). {\em The end of time}.
           Oxford University Press, New York.

\bibitem{} Barbour, J. B. and Bertotti, B. (1982). Mach's principle
           and the structure of dynamical theories.
           {\em Proc. R. Soc. Lond. A}, {\bf 382}, 295--306.  

\bibitem{} Barbour, J. B. and Foster, B. (2008).
           Constraints and gauge transformations: Dirac's theorem is
           not always valid. arXiv:0808.1223v1 [gr-qc].

\bibitem{} Barbour, J. B., Koslowski, T., and Mercati, F. (2013).
The solution to the problem of time in shape dynamics.
arXiv:1302.6264v1 [gr-qc].

\bibitem{} Barvinsky, A. O. and Kiefer, C. (1998). Wheeler--DeWitt equation
           and Feynman diagrams. {\em Nucl. Phys. B}, {\bf 526},
           509--39.

\bibitem{} Barvinsky, A. O., Kamenshchik, A. Yu., Kiefer, C., and
           Mishakov, I. V. (1999). Decoherence in quantum cosmology
           at the onset of inflation. {\em Nucl. Phys. B}, {\bf 551},
           374--96.

\bibitem{} Barvinsky, A. O., Kamenshchik, A. Yu., Kiefer, C.,
           and Steinwachs, C. (2010).
           Tunneling cosmological state revisited: Origin of inflation
           with a nonminimally coupled standard model Higgs inflaton. 
           {\em Phys. Rev. D}, {\bf 81}, 043530.

\bibitem{} Bassi, A., Lochan, K., Satin, S., Singh, T. P., and
  Ulbricht, H. (2013). Models of wave-function collapse, underlying
  theories, and experimental tests. {\em Rev. Mod. Phys.}, {\bf 85},
  471--527. 

\bibitem{} Bastos, C., Bertolami, O., Costa Dias, N. and Nuno Prata,
  J. (2008). Phase-space noncommutative quantum cosmology.
{\em Phys. Rev. D}, {\bf 78}, 023516.

\bibitem{} Belinskii, V. A., Khalatnikov, I. M., and Lifshitz, E. M. (1982).
           A general solution of the Einstein equations with a time
           singularity. {\em Adv. Phys.}, {\bf 31}, 639--67.

\bibitem{} Bergeron, H., Dapor, A., Gazeau, J. P., and Malkiewicz, P. (2013).
           Wavelet quantum cosmology. arXiv:0653v1 [gr-qc].

\bibitem{} Bern, Z., Carrasco, J. J. M., Dixon, L. J., Johansson, H.,
           and Roiban, R. (2009). Ultraviolet behavior of ${\mathcal
           N}=8$ supergravity at four loops. {\em Phys. Rev. Lett.},
           {\bf 103}, 081301.

\bibitem{} Bertolami, O. and Zarro, C. A. D. (2011). 
          Ho\v{r}ava-Lifshitz quantum cosmology.
          {\em Phys. Rev. D}, {\bf 85}, 044042.

\bibitem{} Bertoni, C., Finelli, F., and Venturi, G. (1996). The
           Born--Oppenheimer approach to the matter-gravity system and 
           unitarity. {\em Class. Quantum Grav.}, {\bf 13}, 2375--83.

\bibitem{} Bini, D., Esposito, G., Kiefer, C., Kr\"amer, M., and Pessina, F.
  (2013). On the modification of the cosmic microwave background
  anisotropy spectrum from canonical quantum gravity. 
 {\em Phys. Rev. D}, {\bf 87}, 104008.

\bibitem{} Birrell, N. D. and Davies, P. C. W. (1982).
           {\em Quantum fields in curved space}. Cambridge University Press,
           Cambridge.

\bibitem{} Bjerrum-Bohr, N. E. J., Donoghue, J. F., and
           Holstein, B. R. (2003). Quantum gravitational corrections
           to the nonrelativistic scattering potential of two masses.
           {\em Phys. Rev. D}, {\bf 67}, 084033.

\bibitem{} Blau, M. and Theisen, S. (2009).
           String theory as a theory of quantum gravity: a status
           report. {\em Gen. Relativ. Gravit.}, {\bf 41}, 743--55.

\bibitem{} Blumenhagen, R., L\"ust, D., and Theisen, S. (2013).
{\em Basic Concepts of String Theory}. Springer, Berlin.

\bibitem{} Bojowald, M. (2011). {\em Quantum cosmology}. 
           Lecture Notes in Physics~835. Springer, Berlin.

\bibitem{} Bojowald, M. (2012). Quantum cosmology: effective theory.
       {\em Class. Quantum Grav.}, {\bf 29}, 213001.

\bibitem{} Bojowald, M., Kiefer, C., and Moniz, P. V. (2010).
           Quantum cosmology for the 21st century: a debate.
           arXiv:1005.2471v1 [gr-qc].

\bibitem{} Bojowald, M., Hoehn, P. A., and Tsobanjan, A. (2011).
       An effective approach to the problem of time.
       {\em Class. Quantum Grav.}, {\bf 28}, 035006.

\bibitem{} Bouhmadi-L\'opez, M., Kiefer, C., Sandh\"ofer, B., and
           Moniz, P. V. (2009).
           Quantum fate of singularities in a dark-energy dominated
           universe. {\em Phys. Rev. D}, {\bf 79}, 124035.

\bibitem{} Brown, J. D. and Kucha\v{r}, K. V. (1995). Dust as a standard
           of space and time in canonical quantum gravity.
           {\em Phys. Rev.~D}, {\bf 51}, 5600--29.

\bibitem{} Calzetta, E. (2012). Chaos, decoherence and quantum
  cosmology. {\em Class. Quantum Grav.}, {\bf 29}, 143001.

\bibitem{} Calzetta, E. and Gonzalez, J. J. (1995). 
          Chaos and semiclassical limit in quantum cosmology.
          {\em Phys. Rev. D}, {\bf 51}, 6821--8.

\bibitem{} Carlip, S. (2008). Is quantum gravity necessary? 
          {\em Class. Quantum Grav.}, {\bf 25}, 154010.

\bibitem{} Carlip, S. (2010). The small scale structure of
           spacetime. arXiv:1009.1136v1 [gr-qc].

\bibitem{} Carr, B. J. (2003). Primordial black holes as a probe
           of cosmology and high energy physics. In {\em Quantum 
           gravity: from theory to experimental search} (ed. D.~Giulini,
           C. Kiefer, and C.~L\"ammerzahl),~pp.~301--21. 
           Lecture Notes in Physics 631. Springer, Berlin.

\bibitem{} Carr, B. J. (ed.) (2007). {\em Universe or multiverse?}
           Cambridge University Press, Cambridge. 

\bibitem{} Conradi, H. D. and Zeh, H. D. (1991). Quantum cosmology as an
           initial value problem. {\em Phys. Lett.~A}, {\bf 154},
           321--6.

\bibitem{} Cornish, N. J. and Shellard, E. P. S. (1998).
           Chaos in quantum cosmology. {\em Phys. Rev. Lett.}, {\bf
             81}, 3571--4.

\bibitem{} Coule, D. H. (2005). Quantum cosmological models.
           {\em Class. Quantum Grav.}, {\bf 22}, R125--66. 

\bibitem{} D\c{a}browski, M. P., Kiefer, C., and Sandh\"ofer,
  B. (2006). Quantum phantom cosmology. {\em Phys. Rev. D}, {\bf 74},
  044022.

\bibitem{} DeWitt, B. S. (1967). Quantum theory of gravity. I.
           The canonical theory. {\em Phys. Rev.}, {\bf 160},
           1113--48.

\bibitem{} DeWitt, C. M. and Rickles, D. (eds.) (2011). {\em The Role
    of Gravitation in Physics}. Report from the 1957 Chapel Hill
  Conference. Edition Open Access, Berlin. 

\bibitem{} Douglas, M. R. (2003). The statistics of string/M theory vacua.
           {\em J. High Energy Phys.}, 05(2003), 046.

\bibitem{} Everett, H. (1957). `Relative state' formulation of
           quantum mechanics. {\em Rev. Mod. Phys.}, {\bf 29},
           454--62.

\bibitem{} Freese, K., Hill, C. T., and Mueller, M. T. (1985).       	
Covariant functional Schrodinger formalism and application to the
Hawking effect. {\em Nucl. Phys. B}, {\bf 255}, 693--716.  

\bibitem{} Gambini, R. and Pullin, J. (2011).
           {\em A first course in loop quantum gravity}.
           Oxford University Press, Oxford.

\bibitem{} Gasperini, M. and Veneziano, G. (2003). The pre-big bang
           scenario in string cosmology. {\em Phys. Rep.}, {\bf 373},
           1--212.

\bibitem{} Gibbons, G. W. and Hawking, S. W. (1977).  Cosmological
  event horizons, thermodynamics, and particle creation. {\em
    Phys. Rev. D}, {\bf 15}, 2738--51. 

\bibitem{} Giulini, D. and Gro\ss ardt, A. (2011). 
          Gravitationally induced inhibitions of dispersion according
          to the Schr\"odinger--Newton equation. 
          {\em Class. Quantum Grav.}, {\bf 28}, 195026.

\bibitem{} Goroff, M. H. and Sagnotti, A. (1985). Quantum gravity at two loops.
           {\em Phys. Lett. B}, {\bf 160}, 81--6.

\bibitem{} Gronwald, F. and Hehl, F. W. (1996).  On the gauge aspects
           of gravity. arXiv:gr-qc/9602013v1. 

\bibitem{} Halliwell, J. J. (1991). Introductory lectures on quantum cosmology.
           In {\em Quantum cosmology and baby universes} (ed. S. Coleman,
           J. B. Hartle, T. Piran, and S. Weinberg), pp.~159--243.
           World Scientific, Singapore.

\bibitem{} Halliwell, J. J. and Hawking, S. W. (1985). 
           Origin of structure in the
           Universe. {\em Phys. Rev. D}, {\bf 31}, 1777--91.

\bibitem{} Halliwell, J. J. and Louko, J. (1990). 
           Steepest descent contours in the path integral approach 
    to quantum cosmology. 3. A general method with applications to anisotropic
    minisuperspace models. {\em Phys. Rev. D}, {\bf 42}, 3997--4031.


\bibitem{} Hamber, H. W. (2009). {\em Quantum gravitation -- The Feynman
       path integral approach}. Springer, Berlin.

\bibitem{} Hartle, J. B. and Hawking, S. W. (1983). Wave function of the
           Universe. {\em Phys. Rev.~D}, {\bf 28}, 2960--75.

\bibitem{} Hartle, J. B., Hawking, S. W., and Hertog, T. (2011).
 Local observation in eternal inflation. {\em Phys. Rev. Lett.}, {\bf
   106}, 141302.

\bibitem{} Hawking, S. W. (1975). Particle creation by black holes.
           {\em Commun. Math. Phys.}, {\bf 43}, 199--220; Erratum {\em
           ibid.} {\bf 46}, 206 (1976).

\bibitem{} Hawking, S. W. (1976). Breakdown of predictability in
  gravitational collapse. {\em Phys. Rev.~D}, {\bf 14}, 2460--73. 

\bibitem{} Hawking, S. W. (1982). The boundary conditions of the universe.
           {\em Pontificia Academiae Scientarium Scripta Varia},
           {\bf 48}, 563--74.

\bibitem{} Hawking, S. W. (1984). The quantum state of the universe.
           {\em Nucl. Phys. B}, {\bf 239}, 257--76.

\bibitem{} Hawking, S. W. and Penrose, R. (1996). {\em The nature
           of space and time}. Princeton University Press, Princeton.

\bibitem{} Horowitz, G. T. (1998). Quantum states of black holes. 
           In {\em Black holes and relativistic stars} (ed. R. M. Wald),
           pp.~241--66. The University of Chicago Press, Chicago.


\bibitem{} Isham, C. J. (1987). Quantum gravity. In {\em General
           relativity and gravitation}
           (ed. M. A. H. MacCallum), pp.~99--129. Cambridge University
           Press, Cambridge.

\bibitem{} Isham, C. J. (1993). Canonical quantum gravity and the problem of
           time. In {\em Integrable systems, quantum groups, and
           quantum field theory} (ed. L. A. Ibort and M. A. Rodr\'{\i}guez),
           pp.~157--287. Kluwer, Dordrecht.


\bibitem{} Joos, E., Zeh, H. D., Kiefer, C., Giulini, D., Kupsch, J., 
           and Stamatescu, I.-O. (2003). {\em Decoherence and the
           appearance of a classical world in quantum theory}, 2nd
           edn. Springer, Berlin.

\bibitem{} Joshi, P. S. (2013). The rainbows of
  gravity. arXiv:1305.1005v1 [gr-qc].

\bibitem{} Kamenshchik, A. Y., Kiefer, C., and Sandh\"ofer, B. (2007).
           Quantum cosmology with a big-brake singularity.
           {\em Phys. Rev. D}, {\bf 76}, 064032.

\bibitem{} Kamenshchik, A. Y. and Manti, S. (2012). Classical and
  quantum big brake cosmology for scalar field and tachyonic models.
 {\em Phys. Rev. D}, {\bf 85}, 123518.

\bibitem{} Kiefer, C. (1987). Continuous measurement of minisuperspace
           variables by higher multipoles. {\em Class. Quantum Grav.},
           {\bf 4}, 1369--82.

\bibitem{} Kiefer, C. (1988). Wave packets in minisuperspace.
           {\em Phys. Rev. D}, {\bf 38}, 1761--72.

\bibitem{} Kiefer, C. (1991). On the meaning of path integrals in
            quantum cosmology. {\em Ann. Phys. (NY)}, {\bf 207}, 53--70.

\bibitem{} Kiefer, C. (2001). Hawking radiation from
  decoherence. {\em Class. Quantum Grav.}, {\bf 18}, L151--4.

\bibitem{} Kiefer, C. (2004). Is there an information-loss problem
           for black holes? In {\em Decoherence and
           entropy in complex systems} (ed. H.-T. Elze). 
           Lecture Notes in Physics 633. Spring\-er, Berlin.

\bibitem{} Kiefer, C. (2009). Can the arrow of time be understood from
quantum cosmology? arXiv:0910.5836v1 [gr-qc].

\bibitem{} Kiefer, C. (2012). {\em Quantum gravity}, 3rd edn. Oxford
  University Press, Oxford.

\bibitem{} Kiefer, C. and Kolland, G. (2008). 
           Gibbs' paradox and black-hole entropy.
           {\em Gen. Relativ. Gravit.}, {\bf 40}, 1327--39.

\bibitem{} Kiefer, C. and Kr\"amer, M. (2012).
Quantum Gravitational Contributions to the CMB Anisotropy Spectrum.
 {\em Phys. Rev. Lett.}, {\bf 108}, 021301.

\bibitem{} Kiefer, C. and Polarski, D. (2008).
Why do cosmological perturbations look classical to us?.
 arXiv:0810.0087v2 [astro-ph].

\bibitem{} Kiefer, C. and Sandh\"ofer, B. (2008). Quantum Cosmology.
            arXiv:0804.0672v2 [gr-qc]. 

\bibitem{} Kiefer, C. and Schell, C. (2013). Interpretation of the
  triad orientations in loop quantum cosmology. {\em Class. Quantum
    Grav.}, {\bf 30}, 035008.

\bibitem{} Kiefer, C. and Singh, T. P. (1991). Quantum gravitational
           correction terms to the functional Schr\"odinger equation.
           {\em Phys. Rev. D}, {\bf 44}, 1067--76.

\bibitem{} Kiefer, C. and Zeh, H. D. (1995). Arrow of time in a recollapsing
           quantum universe. {\em Phys. Rev. D}, {\bf 51}, 4145--53.

\bibitem{} Kiefer, C., Lohmar, I., Polarski, D., and Starobinsky,
A. A. (2007). Pointer states for primordial fluctuations in
inflationary cosmology. {\em Class. Quantum Grav.}, {\bf 24}, 1699--1718.

\bibitem{} Kiefer, C., Marto, J., and Moniz, P. V. (2009).
           Indefinite oscillators and black-hole evaporation.
           {\em Annalen der Physik}, 8th series, {\bf 18}, 722--35.

\bibitem{} Kleinschmidt, A., Koehn, M., and Nicolai, H. (2009).
           Supersymmetric quantum cosmological billiards.
           {\em Phys. Rev. D}, {\bf 80}, 061701(R).

\bibitem{} Kim, S. P. (2013). Massive scalar field quantum cosmology.
         arXiv:1304.7439v1 [gr-qc].

\bibitem{} Klauder, J. R. (2010). An affinity for affine quantum
gravity. arXiv:1003.2617v1 [gr-qc].

\bibitem{} Kucha\v{r}, K. V. (1992). Time and interpretations of quantum
           gravity. In {\em Proc. 4th Canadian Conf.
           General relativity Relativistic Astrophysics}
           (ed. G. Kunstatter, D. Vincent, and J. Williams), pp.~211--314.
           World Scientific, Singapore. 

\bibitem{} Landau, S., Sc\'occola, C. G., and Sudarsky, D. (2012).
Cosmological constraints on nonstandard inflationary quantum collapse
models. {\em Phys. Rev. D}, {\bf 85}, 123001.

\bibitem{} MacGibbon, J. H. (1991). Quark- and
gluon-jet emission from primordial black holes. II. The emission over
the black-hole lifetime. {\em Phys. Rev. D}, {\bf 44}, 376--92.

\bibitem{} Maldacena, J. (2005). The illusion of gravity.
         {\em Scientific American}, November 2005, 56--63.

\bibitem{} Maldacena, J. (2011). The gauge/gravity duality.
           arXiv:1106.6073v1 [hep-th].

\bibitem{} Misner, C. W., Thorne, K. S., and Wheeler, J. A. (1973).
           {\em Gravitation}. Freeman, San Francisco.

\bibitem{} Moniz, P. V. (2010). {\em Quantum cosmology -- the
supersymmetric perspective}, Vols.~1 and~2. Lecture Notes in
Physics~804. Springer, Berlin.

\bibitem{} Montani, G., Battisti, M. V., Benini, R., and Imponente,
  G. (2011). {\em Primordial Cosmology}. World Scientific, Singapore.

\bibitem{} Montesinos, M., Rovelli, C., and Thiemann, T. (1999). 
SL(2,R) model with two Hamiltonian constraints. {\em Phys. Rev. D},
{\bf 60}, 044009.

\bibitem{} Montesinos, M. and Rovelli, C. (2001). Statistical
  mechanics of generally covariant quantum theories: a Boltzmann-like
  approach. {\em Class. Quantum Grav.}, {\bf 18}, 555--69.

\bibitem{} Nicolai, H. (2013). Quantum Gravity: the view from particle
           physics. arXiv:1301.5481v1 [gr-qc].

\bibitem{} Nink, A. and Reuter, M. (2012). On quantum gravity,
  asymptotic safety, and paramagnetic dominance. arXiv:1212.4325v1 [hep-th].

\bibitem{} Oriti, D. (ed.) (2009). {\em Approaches to quantum
           gravity}. Cambridge University Press, Cambridge. 

\bibitem{} Padmanabhan, T. (2010). Thermodynamical aspects of gravity:
new insights. {\em Rep. Prog. Phys.}, {\bf 73}, 046901.

\bibitem{} Page, D. N. (1994). Black hole information. In {\em
    Proceedings of the 5th Canadian conference on general relativity
    and relativistic astrophysics} (ed. R. Mann and R. McLenaghan),
  pp.~1--41. World Scientific, Singapore. 

\bibitem{} Page, D. N. (2013). Time dependence of Hawking radiation
  entropy. arXiv:1301.4995v1 [hep-th].

\bibitem{} Page, D. N. and Geilker, C. D. (1981). Indirect evidence
           for quantum gravity. {\em Phys. Rev. Lett.}, {\bf 47},
           979--82.

\bibitem{} Penrose, R. (1981). Time-asymmetry and quantum gravity.
           In {\em Quantum gravity}, Vol.~2 (ed. C. J. Isham,
           R.~Penrose, and D. W.~Sciama),~pp.~242--72. Clarendon Press, Oxford.

\bibitem{} Penrose, R. (1996). On gravity's role in quantum state reduction.
           {\em Gen. Rel. Grav.}, {\bf 28}, 581--600.

\bibitem{} Penrose, R. (2009).  	
Black holes, quantum theory and cosmology.
{\em J.Phys.Conf.Ser.}, {\bf 174}, 012001.

\bibitem{} Pinto-Neto, N., Falciano, F. T., Pereira, R., and Sergio
  Santini, E. (2012). Wheeler--DeWitt quantization can solve the
  singularity problem. {\em Phys. Rev. D}, {\bf 86}, 063504.

\bibitem{} Pons, J. M., Salisbury, D. C., and Sundermeyer,
K. A. (2009).
Revisiting observables in generally covariant theories in the light of
gauge fixing methods. {\em Phys. Rev. D}, {\bf 80}, 084015.

\bibitem{} Rendall, A. D. (2005). The nature of spacetime
  singularities. arXiv: gr-qc/0503112v1. 

\bibitem{} Rovelli, C. (1991). Time in quantum gravity: An hypothesis.
 {\em Phys. Rev. D}, {\bf 43}, 442--56.

\bibitem{} Rovelli, C. (1993). Statistical mechanics of gravity and
  the thermodynamical origin of time. {\em Class. Quantum Grav.}, {\bf
    10},  1549--66. 

\bibitem{} Rovelli, C. (2004). {\em Quantum gravity}. Cambridge University
           Press, Cambridge.

\bibitem{} Rovelli, C. (2013). Covariant loop gravity. In
{\em Quantum Gravity and Quantum Cosmology} (ed.
 G. Calcagni, L. Papantonopoulos, G. Siopsis, and
  N. Tsamis), pp. 57--66.
Lecture Notes in Physics 863. Springer, Berlin.

\bibitem{} Singh, T. P. (2005). Quantum mechanics without
  spacetime: a case for noncommutative geometry.
    arXiv:gr-qc/0510042v1.

\bibitem{} Strominger, A. (2009). Five problems in quantum gravity.
           {\em Nucl. Phys. Proc. Suppl.}, {\bf 192--193}, 119--25.

\bibitem{} Strominger, A. and Vafa, C. (1996). Microscopic origin of the
           Bekenstein--Hawking entropy. {\em Phys. Lett.~B}, {\bf 379},
           99--104.

\bibitem{} Thiemann, T. (2007). {\em Modern canonical 
           quantum general relativity}. Cambridge University Press,
           Cambridge. 

\bibitem{} Thirolf, P. G. {\em et al}. (2009). Signatures of the Unruh
   effect via high-power, short-pulse lasers. {\em Eur. Phys. J. D}, {\bf
   55}, 379--89.

\bibitem{} Torre, C. G. (1993). Is general relativity an 
             ``already parametrized'' theory? {\em Phys. Rev. D}, {\bf 46},
             3231--4.

\bibitem{} Unruh, W. G. (1976). Notes on black-hole evaporation.
           {\em Phys. Rev. D}, {\bf 14}, 870--92. 

\bibitem{} Vaz, C., Gutti, S., Kiefer, C., Singh, T. P., and
           Wijewardhana, L. C. R. (2008).
           Mass spectrum and statistical entropy of the BTZ black hole
           from canonical quantum gravity. 
           {\em Phys. Rev. D}, {\bf 77}, 064021.

\bibitem{} Vilenkin, A. (1989). Interpretation of the wave function of the
           Universe. {\em Phys. Rev. D}, {\bf 39}, 1116--22. 

\bibitem{} Vilenkin, A. (2003). Quantum cosmology and eternal inflation.
           In {\em The future of theoretical physics and cosmology}
           (ed. G. W. Gibbons, E. P. S. Shellard, and S. J. Rankin),
           pp.~649--66. Cambridge University Press, Cambridge.

\bibitem{} Vilenkin, A. (2013{\em a}). Arrows of time and the beginning of
  the universe. arXiv:1305.3836v1 [hep-th].

\bibitem{} Vilenkin, A. (2013{\em b}). Global structure of the
  multiverse and the measure problem.  arXiv:1301.0121v2 [hep-th].

\bibitem{} Wheeler, J. A. (1968). Superspace and the nature of quantum
           geometrodynamics. In {\em Battelle rencontres} (ed. C. M.
           DeWitt and J. A. Wheeler), pp.~242--307. Benjamin, New
           York.

\bibitem{} Weinberg, S. (1995). {\em The quantum theory of fields,
           Vol.~I (Foundations)}. Cambridge University Press, Cambridge.
 
\bibitem{} Wiltshire, D. L. (1996). An introduction to quantum
  cosmology. In {\em Cosmology: the physics of the Universe}
  (ed. B. Robson, N. Visvanathan, and W.S. Woolcock), pp.~473--531.
  World Scientific, Singapore. For a related version, see
  arXiv:\-gr-qc/0101003v2.     

\bibitem{} Zeh, H. D. (1986). Emergence of classical time from a
           universal wave function. {\em Phys. Lett. A}, {\bf 116}, 9--12.  

\bibitem{} Zeh, H. D. (1988). Time in quantum gravity. {\em Phys. Lett. A},
           {\bf 126}, 311--7.

\bibitem{} Zeh, H. D. (2005). Where has all the information gone?
           {\em Phys. Lett. A}, {\bf 347}, 1--7.

\bibitem{} Zeh, H. D. (2007). {\em The physical basis of the direction
           of time}, 5th edn. Springer, Berlin.

\bibitem{} Zeh, H. D. (2011). Feynman's quantum theory.
           {\em Eur. Phys. J. H}, {\bf 36}, 147--58.

\bibitem{} Zurek, W. H. and Paz, J. P. (1995). Quantum chaos: a
  decoherent definition. {\em Physica D}, {\bf 83}, 300--8.

\bibitem{} Zych, M.,  Costa, F.,  Pikovski, I.,  Ralph, T. C., and
  Brukner, C. (2012). General relativistic effects in quantum interference of
photons. {\em Class. Quantum Grav.}, {\bf 29}, 224010.

\endthebibliography
\end{document}